\documentclass{article}
\usepackage[totalwidth=480pt, totalheight=680pt]{geometry}
\usepackage{amsfonts}
\usepackage{amsmath}
\usepackage{subfigure}
\usepackage{graphicx}

\begin{document}

\newcommand{\hi}{H\,{\sc i}}
\newcommand{\civ}{C\,{\sc iv}}
\newcommand{\cii}{C\,{\sc ii}}
\newcommand{\lya}{Ly$\alpha$}
\newcommand{\mgii}{Mg\,{\sc ii}}
\newcommand{\oii}{O\,{\sc ii}}
\newcommand{\feii}{Fe\,{\sc ii}}
\newcommand{\asec}{$^{\prime\prime}$}
\newcommand{\lam}{$\lambda$}
\newcommand{\lamlam}{$\lambda\lambda$}
\newcommand{\lamlamlam}{$\lambda\lambda\lambda$}
\newcommand{\nodata}{...}
\newcommand{\aj}{Astron.~J.}
\newcommand{\actaa}{Acta Astron.}
\newcommand{\araa}{Annu. Rev. Astron. Astrophys.}
\newcommand{\apj}{Astrophys.~J.}
\newcommand{\apjl}{Astrophys. J. Lett.}
\newcommand{\apjs}{Astrophys. J. Suppl. Ser.}
\newcommand{\ao}{Appl. Opt.}
\newcommand{\apss}{Astrophys. Space Sci.}
\newcommand{\aap}{Astron. Astrophys.}
\newcommand{\aapr}{Astron. Astrophys. Rev.}
\newcommand{\aaps}{Astron. Astrophys. Suppl. Ser.}
\newcommand{\azh}{Astron. Zh. }
\newcommand{\baas}{Bull. Am. Astron. Soc.}
\newcommand{\caa}{Chin. Astron. Astrophys.}
\newcommand{\cjaa}{Chin. J. Astron. Astrophys.}
\newcommand{\icarus}{Icarus}
\newcommand{\jcap}{J. Cosmol. Astropart. Phys.}
\newcommand{\jrasc}{J.~R. Astron. Soc. Can.}
\newcommand{\memras}{Mem. R. Astron. Soc.}
\newcommand{\mnras}{Mon. Not. R. Astron. Soc.}
\newcommand{\na}{New Astron.}
\newcommand{\nar}{New Astron. Rev.}
\newcommand{\pra}{Phys. Rev.~A}
\newcommand{\prb}{Phys. Rev.~B}
\newcommand{\prc}{Phys. Rev.~C}
\newcommand{\prd}{Phys. Rev.~D}
\newcommand{\pre}{Phys. Rev.~E}
\newcommand{\prl}{Phys. Rev. Lett.}
\newcommand{\pasa}{Proc. Astron. Soc. Aust.}
\newcommand{\pasp}{Publ. Astron. Soc. Pac.}
\newcommand{\pasj}{Publ. Astron. Soc. Jpn.}
\newcommand{\qjras}{Q. J. R. Astron. Soc.}
\newcommand{\rmxaa}{Rev. Mexicana Astron. Astrofis.}
\newcommand{\skytel}{Sky Telesc.}
\newcommand{\solphys}{Sol. Phys.}
\newcommand{\sovast}{Soviet Astron.}
\newcommand{\ssr}{Space Sci. Rev.}
\newcommand{\zap}{Z. Astrophys.}
\newcommand{\nat}{Nature}
\newcommand{\iaucirc}{IAU Circ.}
\newcommand{\aplett}{Astrophys. Lett.}
\newcommand{\apspr}{Astrophys. Space Phys. Res.}
\newcommand{\bain}{Bull. Astron. Inst. Neth.}
\newcommand{\fcp}{Fundam. Cosmic Phys.}
\newcommand{\gca}{Geochim. Cosmochim. Acta}
\newcommand{\grl}{Geophys. Res. Lett.}
\newcommand{\jcp}{J. Chem. Phys.}
\newcommand{\jgr}{J. Geophys. Res.}
\newcommand{\jqsrt}{J. Quant. Spec. Radiat. Transf.}
\newcommand{\memsai}{Mem. Soc. Astron. Italiana}
\newcommand{\nphysa}{Nucl. Phys.~A}
\newcommand{\physrep}{Phys. Rep.}
\newcommand{\physscr}{Phys. Scr.}
\newcommand{\planss}{Planet. Space Sci.}
\newcommand{\procspie}{Proc. SPIE}

\newcommand{\farcs}{\mbox{\ensuremath{.\!\!^{\prime\prime}}}}%

\title{Investigating Mg\,{\sc ii} Absorption in Paired Quasar Sight-Lines}
\author{J.A.~Rogerson, P.B.~Hall}

\date{Submitted 17 May 2011, Accepted 1 December 2011}


\maketitle

\begin{abstract}
We test whether the Tinker \&\ Chen model of \mgii\ absorption due to the gaseous halo around a galaxy can reproduce absorption in quasar pairs (both lensed and physical) and lensed triples and quads from the literature.  These quasars exhibit absorption from a total of 38 \mgii\ systems spanning $z=0.043 - 2.066$ with mean redshift $\langle z \rangle=1.099$ and weighted mean rest-frame equivalent width of 0.87 \AA.  Using the Tinker \&\ Chen model to generate simulated sight-lines, we marginalize the unknown parameters of the absorbing galaxies: dark matter halo mass, impact parameter, and azimuthal angle on the sky.  We determine the ability of the model to statistically reproduce the observed variation in \mgii\ absorption strength between paired sight-lines for different values of the gas covering fraction $f_c$ and the characteristic length scale $\ell_c$, within which the variation in absorption equivalent widths between sight-lines exponentially decreases.  We find a best-fit $f_c=0.60\pm0.15\%$ and $\ell_c < 8~h^{-1}_{70}~{\rm kpc}$ (1$\sigma$ confidence limits), with smaller $f_c$ allowed at larger $\ell_c$.  At 99.7\%\ confidence, we are able to rule out $f_c>0.87$ for all values of $\ell_c$ and the region where $\ell_c<1.0~h^{-1}_{70}$ kpc and $f_c<0.3$.
\end{abstract}



\section{Introduction}\label{intro}

Intervening metal absorbers found in the spectra of background quasars have been known to be associated with galaxies at intermediate redshifts since the work of \nocite{BB91}{Bergeron} \& {Boiss{\'e}} (1991).  As a result, quasar sight-lines have become a pivotal tool in probing the gaseous halos surrounding these galaxies, using (among other transitions) the \mgii\ $\lambda\lambda2796.352,2803.531$\,\AA\ doublet (e.g., \nocite{2011ApJ...733..105K}{Kacprzak} {et~al.} 2011).

The physical origin of the gaseous halo and its evolving interplay with its host galaxy, however, are not well understood.  Gas in the halo could be fuel for continued star formation in the galaxy (e.g., \nocite{MB04}{Maller} \& {Bullock} 2004), or could be a result of gas blown out by galactic winds (e.g., \nocite{SSP10}{Steidel} {et~al.} 2010), or both.
It is clear that the gas is clumped and not distributed spherically (e.g., \nocite{CKSiau05}{Churchill}, {Kacprzak} \&  {Steidel} 2005), but the typical covering fraction is still uncertain.  The spatial extent of the \mgii\ absorbing region has been constrained by various studies, although they do not always agree.  The extent could be related to the mass of the galaxy (e.g., \nocite{TC08}{Tinker} \& {Chen} 2008, \nocite{SDP94}{Steidel}, {Dickinson} \& {Persson} 1994, \nocite{LS11}{Lovegrove} \& {Simcoe} 2011) or to the colour and star-formation rate of the galaxy (e.g., \nocite{ZMNQ07}{Zibetti} {et~al.} 2007, \nocite{MW09}{M{\'e}nard} {et~al.} 2009 [but see \nocite{LC11}{L{\'o}pez} \& {Chen} 2011], \nocite{SSP10}{Steidel} {et~al.} 2010).  

The development of new observational and measurement techniques as well as new models of \mgii\ absorption is required to piece together the picture of the gaseous halo.


\nocite{TC08}{Tinker} \& {Chen} (2008) (hereafter TC08) have constructed a semi-analytical model for gas absorbing in the \mgii~\lam2796~\AA\ transition within a dark matter (DM) halo using an isothermal density profile.  The rest-frame absorption equivalent width $EW(2796)$ (absorption due to the 2796~\AA\ transition of the \mgii\ doublet at some projected distance $r$ from the centre of the absorbing gaseous halo, hereafter $EW$) is determined by the amount of cold gas, in the form of discrete clouds, along the line of sight (LOS) through a halo.  They constrained their model using a \mgii-LRG cross-correlation function and a \mgii\ absorbing frequency distribution function.

In \nocite{CT08}{Chen} \& {Tinker} (2008), the authors further constrained the TC08 model using a sample of \mgii\ absorption features near known galaxies.  The sample consisted of 13 galaxy and absorber pairs and 10 galaxies that do not produce \mgii\ absorption lines to within sensitive upper limits.  
A maximum likelihood analysis showed that the model best describes the observations if the covering fraction is 80-86\%.  The data had a weighted mean rest-frame equivalent width of 0.34~\AA.

In this study we present a new test of the TC08 model, using \mgii\ absorption features in the lines of sight to lensed and multiple quasar systems.  These systems can provide two or more lines of sight through the same galaxy halo.  By comparing the coincident and anti-coincident absorption systems in multiple lines of sight, constraints on both the covering fraction and coherence scale can be determined (e.g., \nocite{EIP04}{Ellison} {et~al.} 2004).  Multiple quasar systems may also be useful in measuring the azimuthal profiles of \mgii\ absorbers around galaxies, a topic studied statistically by \nocite{BLK11}{Bordoloi} {et~al.} (2011).  

We have compiled a non-exhaustive list from the literature of quasar pairs, triples, and quads (hereafter referred to as asterisms) where \mgii\ absorption is observed in one or all lines of sight.  Using a statistical approach, we ascertain how well the TC08 model can reproduce the observed variation in \mgii\ absorption between sight-lines for various values of the TC08 model parameters.

We describe our literature sample in \S\ \ref{data}.  Descriptions of the theoretical framework of the TC08 model and the method by which we generate our simulations are given in \S\ \ref{framework}.  Our analysis and results are described in \S\ \ref{analysis}.  We discuss our results in \S\ \ref{discussion}, and summarize our study in \S\ \ref{summary}.

In this paper, we adopt $H_0=70$ km s$^{-1}$ Mpc$^{-1}$, $h_{70}=H_0/70$ km s$^{-1}$ Mpc$^{-1}$, $\Omega_M=0.3$, $\Omega_{\Lambda}=0.7$ \nocite{wmap3}({Spergel} {et~al.} 2007).


\section{Data}\label{data}

We have tabulated data on the following quasars, for which references are given in Tables \ref{pairstable} and \ref{quadEWtable}: 7 pairs, both physical and lensed, (Q1343+266AB; HE 1104-1805AB; HS 1216+5032AB; Q0957+561AB; SDSS J1029+2623BC; SDSS J0904+1512AB; and SDSS J1054+2733AB), 2 triply lensed quasars (Q2237+0305ABC and APM08279+5255ABC), and 2 quadruply lensed quasars (H1413+1143ABCD and SDSS J1004+4112ABCD).  SDSS J1029+2632 B and C are part of a triply lensed source quasar, however, only these two images had available spectra and so we treat it as if it were a lensed pair.  Q2237+0305 is a quadruply lensed source, but only images A, B, and C had available spectra, so we consider it a triple.  Also of note in Q2237+0305, the absorption system at $z=0.827$ only has data for images A and B, therefore we include that one system in the pairs data set.  

In the cases of SDSS J1004+4112ABCD, SDSS J1029+2623BC, SDSS J0904+1512AB, and SDSS J1054+2733AB we measured the EW ourselves using spectra provided by M. Oguri.  For Q2237+0305, we obtained the EW by converting the column densities given in Rauch et al. (2002).

The absorption systems at $z=0.04299~{\rm and}~0.04431$ found in the physical pair HS 1216+5032 have been included in this set despite their questionable nature (see \S\ 4.2 in \nocite{LH00}{Lopez}, {Hagen} \& {Reimers} 2000).  By not including the two features in our data set, no major changes to the results were observed.

Our sample harbours 38 unique absorption features, wherein at least one of the lines of sight in the asterism has a rest-frame equivalent width measurement of the \mgii~\lam2796~\AA\ transition.  The other line(s) of sight in each asterism may not have any detectable \mgii\ absorption, in which case they are assigned a $3\sigma$ upper limit to their EW (using the RMS noise in the data).\footnote{The non-detection upper limits were measured in the same manner as the detections in each system.}  All data is summarized in Table \ref{pairstable} (pairs) and Table \ref{quadEWtable} (triples and quads) in order of date published (references are found in last column of pairs table, first column of triples/quads table).  Table \ref{pairstable} lists the object name and components, emission redshift, the lens redshift (if applicable), the absorber redshift, the proper separation between the quasar images at the absorber redshift ($h^{-1}_{70}$~kpc), and the EW of both lines of sight (measured in \AA).  Table \ref{quadEWtable} lists similar values, but lists the maximum and minimum separation in the object, and does not list the individual proper separations.  The weighted mean $EW$ in our data set is 0.87 \AA.

Our data sample excludes objects with line of sight separations much larger than the scale size of gaseous halos described by the TC08 model (proper separations greater than $\sim$300 kpc).

\begin{figure}
\includegraphics[width=150mm]{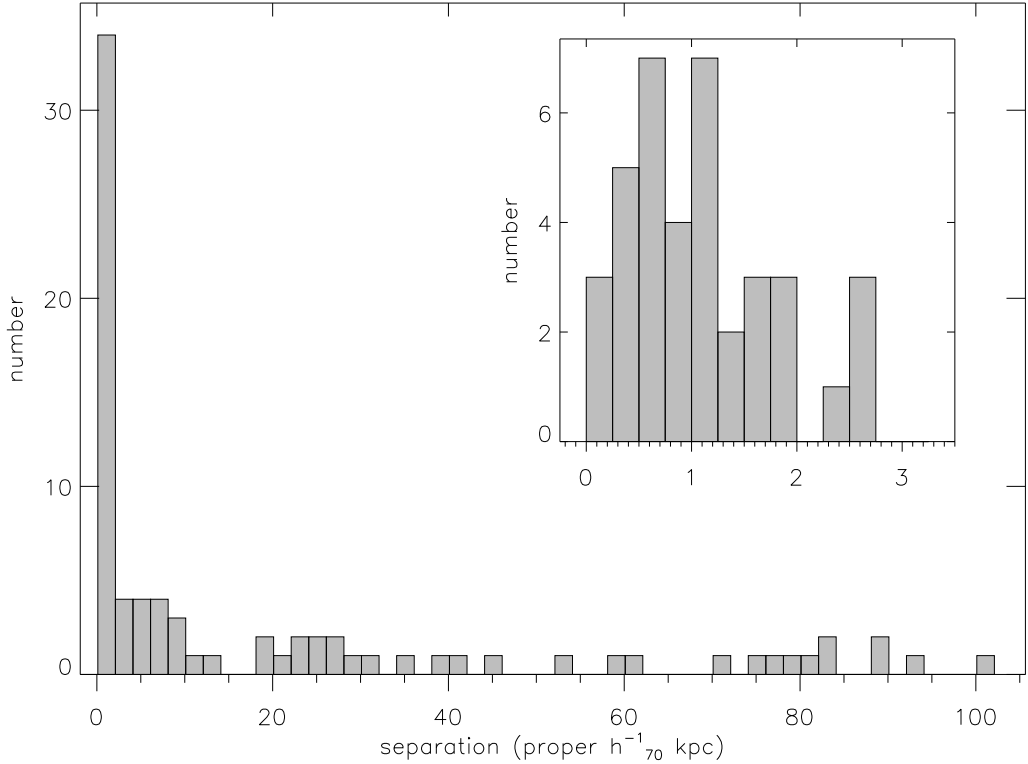}
\caption{Histogram of all separations in the data set. Bins are 0.25 proper $h^{-1}_{70}$ kpc in the inset and 2.0 proper $h^{-1}_{70}$ kpc otherwise.  This figure includes all configurations of quads and triples.}
\label{sep_all}
\end{figure}

\begin{table*}
\begin{minipage}{150mm}
\caption{Tabulated quasar pairs from the literature that exhibit \mgii\ absorption at $\lambda2796$~\AA.  See last column for reference.  Separations $d$ are quoted in $h^{-1}_{70}$~kpc in the proper frame.}
\label{pairstable}

\begin{tabular}{@{}l l l l l l}

\hline\hline
 Object & z$_{abs}$ & $d$ & A~(\AA) & B~(\AA) & z$_{source}$, z$_{lens}$, Reference \\
\hline
Q1343+266 & 0.516 & 58.95 & 0.65$\pm$0.07 & $<$0.2 & 2.03, not a lens, Crotts et al. 1994 \\
$\theta=$9\farcs5 &  & & & & \\
\hline
 HE 1104-1805 & 0.5168 & 19.89 & 0.21$\pm$0.02 & $<$0.23 & 2.31, 0.729$^b$, Smette et al. 1995 \\
$\theta=$3\farcs2  & 0.7283 & 23.24 & 0.62$\pm$0.02 & $<$0.27 & \\
 & 1.2799 & 9.75 & $<$0.05 & 0.35$\pm$0.06 & \\
 & 1.3207 & 9.10 & 0.71$\pm$0.02 & 0.35$\pm$0.10 & \\
 & 1.6616 & 4.76 & 0.97$\pm$0.05 & 0.59$\pm$0.11 & \\
\hline
 HS 1216+5032 & 0.04299$^a$ & 7.71 & $<$0.30 & 1.49$\pm$0.54 & 1.45, not a lens, Lopez et al. 2000 \\
  $\theta=$9\farcs1 & 0.04431$^a$ & 7.93 & $<$0.30 & 2.64$\pm$0.92 & \\
  & 0.13542 & 21.83 & $<$0.24 & 0.45$\pm$0.19 & \\
\hline
 Q0957+561 & 1.3911 & 0.235 & 2.27$\pm$0.01 & 2.15$\pm$0.01 & 1.41, 0.361, Churchill et al. 2003 \\
 $\theta=$6\farcs2 & & & & & \\
\hline
 SDSS J1029+2632 B-C & 0.674 & 10.53 & 1.51$\pm$0.07 & 1.45$\pm$0.24 & 2.197, 0.6, Oguri et al. 2008 \\
  $\theta=$1\farcs768 & 1.761 & 1.34 & 0.19$\pm$0.04 & $<$0.4 & \\
  & 1.910 & 0.81 & 0.63$\pm$0.10 & $<$0.6 & \\
\hline
  J0904+1512 & 1.2168 & 0.52 & 2.0$\pm$0.4 & 1.3$\pm$0.4 & 1.826, 0.19, Kayo et al. 2009\\
  $\theta=$1\farcs128 & 1.6127 & 0.14 & 0.3$\pm$0.1 & $<$1.1 & \\
  & 1.6528 & 0.11 & 0.3$\pm$0.1 & $<$0.9 & \\
\hline
  J1054+2733 & 0.6794 & 1.87 & 0.77$\pm$0.1 & 1.07$\pm$0.19 & 1.452, 0.23, Kayo et al. 2009 \\
  $\theta=$1\farcs269 & & & & & \\
\hline
\end{tabular}
\medskip
\\
$a$ See the discussion in \S\ 2. \\
$b$ The lens redshift was determined by \nocite{LC00}{Lidman} {et~al.} (2000).
\end{minipage}

\end{table*}

\begin{table*}
\begin{minipage}{160mm}
\caption{Tabulated quasar triples and quads from the literature.  The value $d_{max}$/$d_{min}$ indicates the largest/smallest proper separations (in units of $h^{-1}_{70}$~kpc) between quasar images in the triple or quad at the given absorption redshift.  For triples the entry '\nodata' is used in the 'D' column.}
\label{quadEWtable}
\begin{tabular}{c c c c c c c c}

\hline\hline
 Object & z$_{abs}$ & A~(\AA) & B~(\AA) & C~(\AA) & D~(\AA) & $d_{max}$ & $d_{min}$ \\
 \hline
H1413+1143 & 0.6089 & 0.25$\pm$0.09 & 0.19$\pm$0.09 & 0.55$\pm$0.09 & 0.48$\pm$0.08 & 9.1 & 5.1 \\
Monier et al. 1998 & $z_{source}=2.54$ & $z_{lens}=1.0^a$ & & & & & \\
\hline
Q2237+0305 & 0.566 & 0.65$\pm$0.07 & 0.65$\pm$0.06 & 0.78$\pm$0.07 & \nodata & 0.54 & 0.4 \\
           & 0.827 & 0.05$\pm$0.06 & 0.07$\pm$0.01 & \nodata$^b$ & \nodata & 0.3 & 0.3 \\
Rauch et al. 2002 & $z_{source}=1.69$ & $z_{lens}=0.039$ & & & & & \\
\hline
SDSS J1004+4112 & 0.676 & 0.91$\pm$0.03 & $<$0.2 & $<$0.2 & 1.0$\pm$0.1 & 101.6 & 26.3 \\
& 0.726 & $<$0.1 & $<$0.2 & $<$0.3 & 2.6$\pm$0.1 & 93.5 & 74.2 \\
& 0.749 & $<$0.1 & $<$0.1 & $<$0.2 & 1.4$\pm$0.1 & 89.5 & 71.0 \\
& 0.833 & $<$0.1 & 0.8$\pm$0.1 & $<$0.2 & $<$0.4 & 77.2 & 19.7 \\
& 1.022 & $<$0.1 & 0.23$\pm$0.03 & $<$0.1 & $<$0.2 & 52.2 & 13.3 \\
& 1.083 & $<$0.1 & $<$0.2 & $<$0.2 & 1.5$\pm$0.1 & 44.9 & 35.6 \\
& 1.226 & $<$0.1 & $<$0.1 & $<$0.2 & 0.5$\pm$0.1 & 31.5 & 25.0 \\
& 1.258 & $<$0.1 & $<$0.1 & $<$0.2 & 0.9$\pm$0.1 & 28.8 & 22.9 \\
Oguri et al. 2004 & $z_{source}=1.734$ & $z_{lens}=0.68$ & & & & & \\
\hline
APM08279+5255 & 1.181 & 2.57$\pm$0.02 & 3.03$\pm$0.04 & 2.88$\pm$0.04  & \nodata  & 2.70 & 1.07  \\
& 1.209 & 0.05$\pm$0.01 & 0.06$\pm$0.02 & $<$0.03 & \nodata & 2.62 & 1.03 \\
& 1.211 & 0.37$\pm$0.01 & $<$0.03 & 0.16$\pm$0.04 & \nodata & 2.61 & 1.03 \\
& 1.291 & 0.08$\pm$0.01 & 0.03$\pm$0.01 & $<$0.03 & \nodata & 2.39 & 0.95  \\
& 1.444 & 0.04$\pm$0.01 & 0.04$\pm$0.01 & $<$0.03 & \nodata & 2.10 & 0.80 \\
& 1.550 & 0.31$\pm$0.01 & $<$0.02 & 0.29$\pm$0.03 & \nodata & 1.81 & 0.72 \\
& 1.552 & 0.24$\pm$0.01 & $<$0.02 & $<$0.03 & \nodata & 1.81 & 0.71 \\
& 1.813 & 0.80$\pm$0.02 & 0.77$\pm$0.02 & 0.44$\pm$0.03 & \nodata & 1.36 & 0.54 \\
& 2.041 & 0.21$\pm$0.02 & 0.24$\pm$0.02 & 0.22$\pm$0.03 & \nodata & 1.07 & 0.42 \\
& 2.066 & 0.31$\pm$0.04 & 0.45$\pm$0.04 & 0.38$\pm$0.04 & \nodata & 1.04 & 0.41 \\
Ellison et al. 2004 & $z_{source}=3.911$ & $z_{lens}=1.062$ & & & & & \\
\hline
\end{tabular}
\medskip
\\
$a$ The lensing object is yet to beidentified, though it is probably at z$>$1 (see Monier et al. 1998).\\
$b$ Negligible \mgii\ absorption was measured in the third image.  This feature is used as part of the pair sample.
\end{minipage}
\end{table*}

\section{Theoretical Framework and Simulations}\label{framework}

In this analysis we have tested the model developed by \nocite{TC08}{Tinker} \& {Chen} (2008) using observational data from the literature.  The TC08 model is a description of how the rest-frame equivalent width of the \mgii~\lam2796~\AA\ transition changes with respect to the impact parameter between the galaxy and the LOS.
Here we describe the framework of the TC08 model and how we applied the tabulated data from the literature and coding to test it.

\subsection{The TC08 model}\label{TC08framework}
The TC08 model is based on populating a dark matter halo with cold baryonic matter, which is represented as \mgii~\lam2796~\AA\ absorbing gas.  The authors invoke a spherical isothermal density profile to approximate the mean density of the gas as a function of distance from the centre of the dark matter halo:
%
\begin{equation}
\rho_g(r)=f_gG_0(r^2+a_h^2)^{-1}
\label{density_profile}
\end{equation}
where $f_g$ is the gas fraction (the fraction of the total baryonic and dark matter mass, $M_h$, in gas form), $a_h$ is the core radius, and
\begin{equation}
G_0 = \frac{M_h(<R_g)/4\pi}{R_g - a_h \tan^{-1}(R_g/a_h)}
\label{G0}
\end{equation}
is a normalization constant with the arctangent in radians.  Only the mass within $R_g$ will contribute to the density profile.  Due to the apparent clumpy nature of the gaseous halo (see, e.g., \nocite{RS99}{Rauch}, {Sargent} \& {Barlow} 1999), the equivalent width of a \mgii~\lam2796~\AA\ absorption is directly proportional to the number of absorbing components along the LOS.  At radii larger than $R_g$, the density of absorbing clumps drops to zero and no absorption is seen.  Integrating the individual effects of each clump along a line of sight $l$ through the gaseous halo at an impact parameter $s$ relative to the centre of the gas distribution yields
\begin{equation}
\begin{array}{l l}
EW(s|M_h) & = EW_0\left[ \frac{2\sigma_{cl}}{M_{cl}}\int_{0}^{\sqrt{R^2_g - s^2}}\rho_g(\sqrt{s^2+l^2})dl ~\right] \\
 & \\
 & =\frac{EW_0\sigma_{cl}f_g}{M_{cl}} \frac{2G_0}{\sqrt{s^2 + a_h^2}}\tan^{-1} \sqrt{\frac{R^2_g - s^2}{s^2 + a^2_h}}
\end{array}
\label{integral}
\end{equation}
where $\sigma_{cl}$ and $M_{cl}$ are the cross section and mean gas mass of an individual clump of gas, $EW_0$ is the equivalent width per clump, and $EW(s|M_h)$ is the rest frame equivalent width for \mgii~\lam2796~\AA\ at $s$ given the dark matter halo mass $M_h$.  

TC08 note that numerical simulations have shown that stable shocks can develop in the centres of dark matter halos due to high accretion rates at high halo masses.  Above a critical halo mass, the dynamical time-scale for matter to accrete onto the dark matter halo is shorter than the post-shock gas cooling time, resulting in temperatures at, or above, the virial temperature.  The result is a shock-heated region at the centre of the gaseous halo in which the cold \mgii\ absorbing gas can no longer survive.

To incorporate this effect, TC08 introduce a parameter $R_{sh}$ 
which is the radius of the transition from shock-heated gas to colder gas.  At $r<R_{sh}$ the gas has been heated and thus no longer can absorb in the \mgii\ transition effectively.  Thus, equation (\ref{integral}) is broken into different regions.

For the cases where $R_{sh}=0$ or $R_{sh} \le s \le R_g$,
\begin{equation}
EW(s|M_h) = \frac{2 A_{EW} G_0}{\sqrt{s^2 + a_h^2}}\tan^{-1} \sqrt{\frac{R^2_g - s^2}{s^2 + a^2_h}}.
\label{W1}
\end{equation}
If the impact parameter is within the shock-heated region ($s<R_{sh}$), then the effect of the reduced \mgii\ absorbing gas is subtracted from the total:
\begin{equation}
EW(s|M_h) = EW_{(R_{sh}=0)} - 
f_{hot}\frac{2 A_{EW} G_0}{\sqrt{s^2 + a_h^2}}\tan^{-1} \sqrt{\frac{R^2_{sh} - s^2}{s^2 + a^2_h}} 
\label{W2}
\end{equation}
where $f_{hot}=1-f_{cold}$ and $f_{cold}$ is the fraction of gas within the shock-heated region that is still able to absorb in \mgii\ (cold gas).  If the entire gaseous halo has been shock-heated ($R_{sh}=R_g$) then we measure the EW by
\begin{equation}
EW(s|M_h) = f_{cold}A_{EW}\frac{2G_0}{\sqrt{s^2 + a_h^2}}\tan^{-1} \sqrt{\frac{R^2_g - s^2}{s^2 + a^2_h}}
\label{W3}
\end{equation}
as only a fraction $f_{cold}$ of the gas within the halo will absorb at the \mgii\ doublet transitions.  Of course, if $s>R_g$ then $EW(s|M_h)=0$.
%
%

In the above formulae TC08 grouped the parameters $EW_0$, $\sigma_{cl}$,$f_g$, and $M_{cl}$ into the single parameter, $A_{EW}$, which represents the mean absorption equivalent width per unit total surface mass density of the cold gas:
\begin{equation}
A_{EW} = \frac{EW_0}{M_{cl}/f_g\sigma_{cl}}.
\label{Aw}
\end{equation}
%
%
%
%

\begin{figure}
\centering
\subfigure{
  \includegraphics[width=84mm]{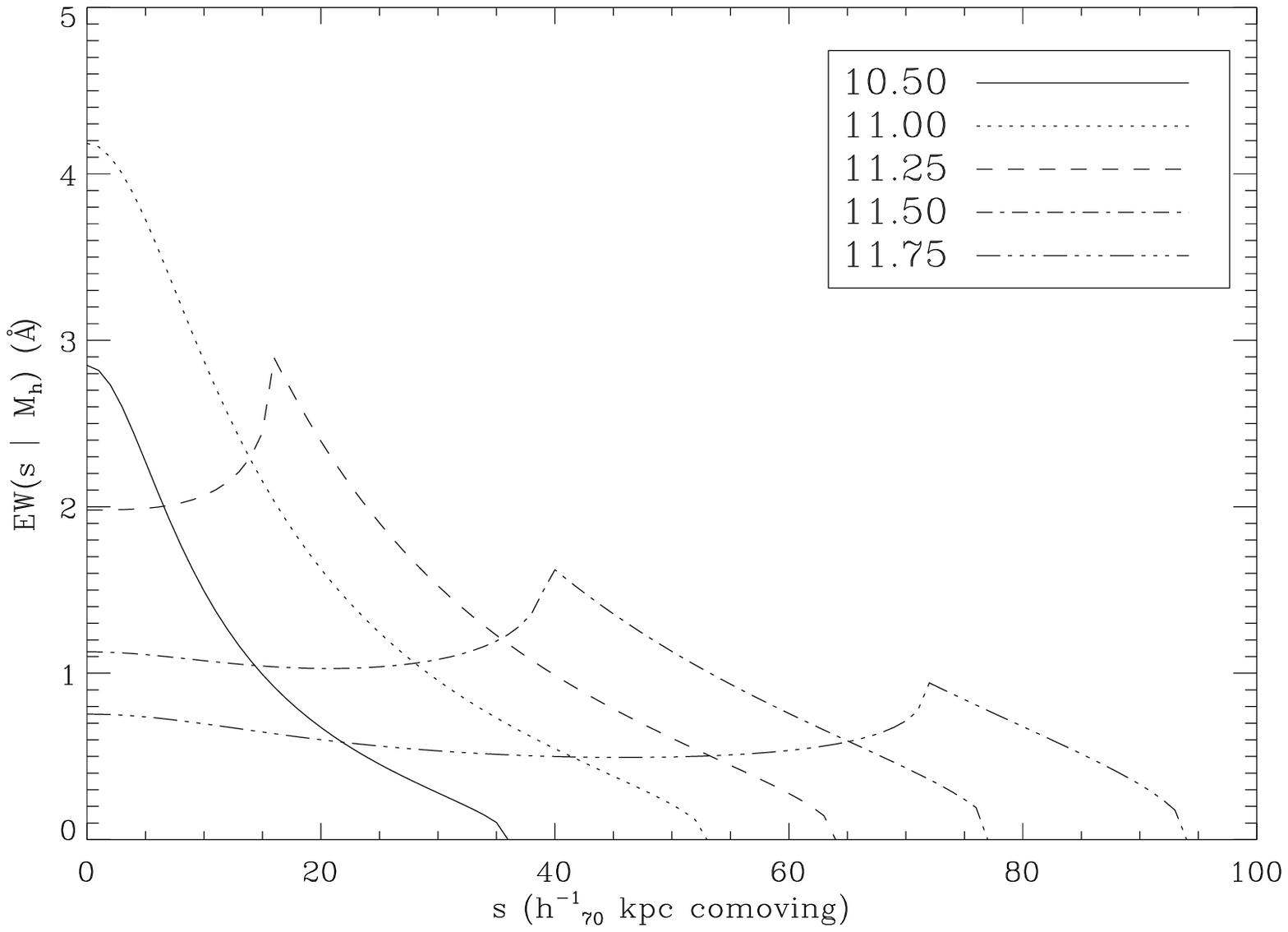}
}
\subfigure{
  \includegraphics[width=84mm]{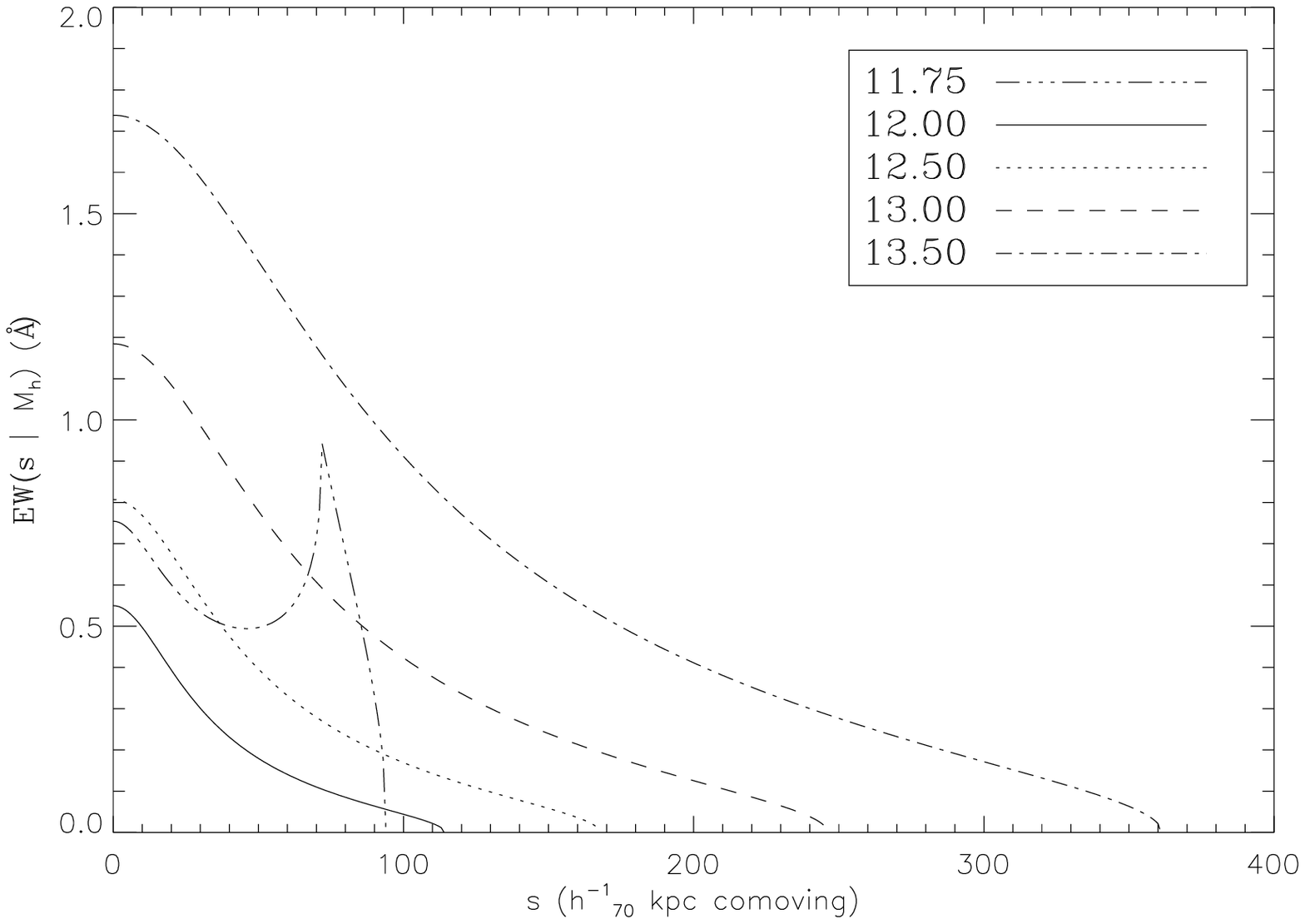}
}
\caption{The equivalent width of the \mgii\ 2796\AA\ transition as a function of impact parameter (in comoving reference frame) predicted by the TC08 model.  Each curve represents a different dark matter halo mass.  The legend provides the curve for each of the values for mass, measured in solar masses and in the log.  In the top figure, the effects of the shock heated region are manifested as a reduction in EW at low impact parameters for masses between $\log M_h/M_{\odot}=11.00$ and $\log M_h/M_{\odot}=11.75$.  At $\log M_h/M_{\odot}=12.00$, the entire gaseous halo becomes shock heated and the underlying form of the density profile is recovered, although, at reduced EW.  In the bottom figure, the model curve for the mass $\log M_h/M_{\odot}=11.75$ is plotted again for reference.}
\label{EWprofile}
\end{figure}

Figure \ref{EWprofile} shows the $EW(s|M_h)$ profile for various values of mass.  Note the cavity created at small impact parameters for halos with shock-heated regions.



The value of $R_g$ (as well as $a_h$ and $R_{sh}$) are dependent on the mass of the halo.  Following TC08 we set 
\begin{equation}
R_g = 80(M_h/10^{12} M_{\odot})^{1/3} h^{-1} {\rm ~kpc~ (comoving)} 
\label{Rg}
\end{equation}
where $h = H_0/(100$ km s$^{-1}$ Mpc$^{-1})$ and $a_h = 0.2R_g$ (TC08 note that the exact choice of $a_h$ has little effect on their results).
Converting to the proper frame and using our Hubble constant, the radius of the gaseous halo is
\begin{equation}
R_g = \frac{114}{1+z_{abs}} \left( \frac{M_h}{10^{12} M_\odot} \right)^{1/3} h_{70}^{-1} {\rm ~kpc}
\label{Rg_new}
\end{equation}
where $h_{70} = H_0/(70$ km s$^{-1}$ Mpc$^{-1})$ and $z_{abs}$ is the redshift of the halo. Hereafter, all coordinates will be in the proper frame.
%
%

TC08 have parameterized the value $R_{sh}$ as
\begin{equation}
\frac{R_{sh}}{R_g} = \hat{R}_{sh}^0 + \gamma_{sh} \log (M_h/10^{12} M_{\odot})
\label{shockradius}
\end{equation}
and forced the bounds $0 \le R_{sh}/R_g \le 1$.  Any values of $R_{sh}/R_g < 0$ correspond to a gaseous halo that has no shock-heated region (i.e., $R_{sh}=0$), and $R_{sh}/R_g > 1$ indicates a gaseous halo that is entirely shock-heated (i.e., $R_{sh}=R_g$).

TC08 utilized a Monte Carlo Markov Chain technique to determine the best fit values to the 4 free parameters in the model.  We adopt the same parameters.  These are  $A_{EW}=534 \: h$ \AA\ cm$^2$ g$^{-1}=374 \: h_{70}$ \AA\ cm$^2$ g$^{-1}$, $f_{cold}=0.061$, $\hat{R}_{sh}^0=1.02$ and $\gamma_{sh}=1.03$.  With these parameters, a shock radius of $R_{sh}=0$ has mass $\log M_h/M_{\odot} \le 11.0 $ and $R_{sh}=R_g$ has mass $\log M_h/M_{\odot} \ge 12.0 $.  The mass range over which the shock heated region appears and encompasses the entire gaseous halo is small in comparison to the mass range used in TC08's analysis.

The TC08 model was tested against 13 galaxy/absorber pairs in \nocite{CT08}{Chen} \& {Tinker} (2008) where a cosmic scatter of 0.25 in the log was observed.
With such a large cosmic scatter, a pair of sight-lines with increasingly smaller separation still has the potential to have significantly different absorption equivalent widths.  We expect there to be some scale at small separations over which the absorption equivalent width will not vary between sight-lines.  To account for this we introduce a characteristic length, denoted $\ell_c$.  Below $\ell_c$, the cosmic scatter is suppressed exponentially so that at zero sight-line separation, the difference in cosmic scatter between sight-lines is zero (see section \ref{simforpairs} for more details).

We include a halo absorbing fraction $f_h(M_h)$ as defined in TC08.  This is the probability that a dark matter halo has any gaseous halo capable of absorbing at all.  More details on this parameter can be found in section \ref{selectmass}.

We also introduce a covering fraction term, $f_c$, which measures the fraction of sight-lines within $R_g$ that do not show absorption (see section \ref{cov_frac} for more details). 

There are a total of seven free parameters required to fully describe a gaseous halo in our model.  TC08 determined the values for five of these parameters.  In this study we have varied the remaining two, the characteristic length and covering fraction, in order to find the best fit values.  The characteristic length $\ell_c$ is varied over $0.1 - 100~h^{-1}_{70}$~kpc and the covering fraction $f_c$ is varied over $0.01 - 1.0$.


\subsection{Simulating Paired Quasar Sight-Lines}\label{application}
We aim to determine if the variation in EW between sight-lines in the observed data set (Tables \ref{pairstable} and \ref{quadEWtable}) is consistent with the TC08 model, and over what range of characteristic length and covering fraction that consistency holds by simulating sight-lines in a gaseous halo described by the TC08 model.

We begin by choosing a mass at random from a halo mass function.  The mass defines the radius of the gaseous halo, and the size of the shock heated region.  We take a random impact parameter from within the gaseous halo and calculate the equivalent width predicted by the TC08 model, based on equations (\ref{W1}-\ref{W3}).  We search through all EW values measured to better than $3\sigma$ in the observed data set for a match to the predicted EW.\footnote{A single predicted EW from the TC08 model can be matched to multiple observed EWs.}  The matched EW represents one LOS from the corresponding quasar asterism.  We use the TC08 model to then predict the EWs of the other lines of sight in the asterism, given the mass of the halo and the impact parameter that was used in the original match.  
For each line of sight in the observed data set with an EW measured to better than $3\sigma$, we compile 900 EWs that match to within observational error.  For each of the 900 matches, we then use the TC08 model to predict the EWs for each of the other lines of sight in the asterism, given an assumed characteristic length scale $\ell_c$ and covering fraction $f_c$ of the absorption.  Using the resulting distribution of predicted EWs, we calculate the probability that the TC08 model can accurately predict the observed variation between lines of sight for each combination of $\ell_c$ and $f_c$.

This statistical approach marginalizes the mass of the dark matter halo, the impact parameter of the sight-line relative to the centre of the gaseous halo, and the orientation of the asterism within the halo.  Below we describe how we determine these parameters, and how we generate simulated samples of EWs to compare with the observed EWs.

\subsubsection{Selecting Mass}\label{selectmass}
We select a mass randomly from a weighted halo mass function.  The weight we have already defined as $f_h(M_h)$, the probability that a given dark matter halo of mass $M_h$ will have a gaseous halo from which absorption can be measured, as not all halos contain a gaseous region.  The halo mass function quantifies the number of dark matter halos as a function of halo mass.  TC08 utilize the halo mass function of \nocite{WAHL06}{Warren} {et~al.} (2006), determined by numerical simulations of structure growth and halo formation.  Multiplying the mass function by $f_h$ results in an absorption weighted halo mass function, allowing proper sampling of the mass range for dark matter halos that would be observed in our data set.

In TC08, however, $f_h$ is degenerate with the covering fraction of the gaseous halo.  So the authors combine the probability that a given dark matter halo will have a gaseous halo that will absorb in \mgii,  and the probability that a given line of sight inside the gaseous region will produce \mgii\ absorption.  In this paper, we are required to explain cases of absorption in only one of two lines of sight that lie within $R_g$.  Since we require there to already be a match to one line of sight in an asterism (discussed in \S\ \ref{simforpairs}), we are not double counting the effects of covering fraction by using $f_h$ and $f_c$.

\begin{figure}
\centering
\subfigure{
 \label{fhprofile}
 \includegraphics[width=84mm]{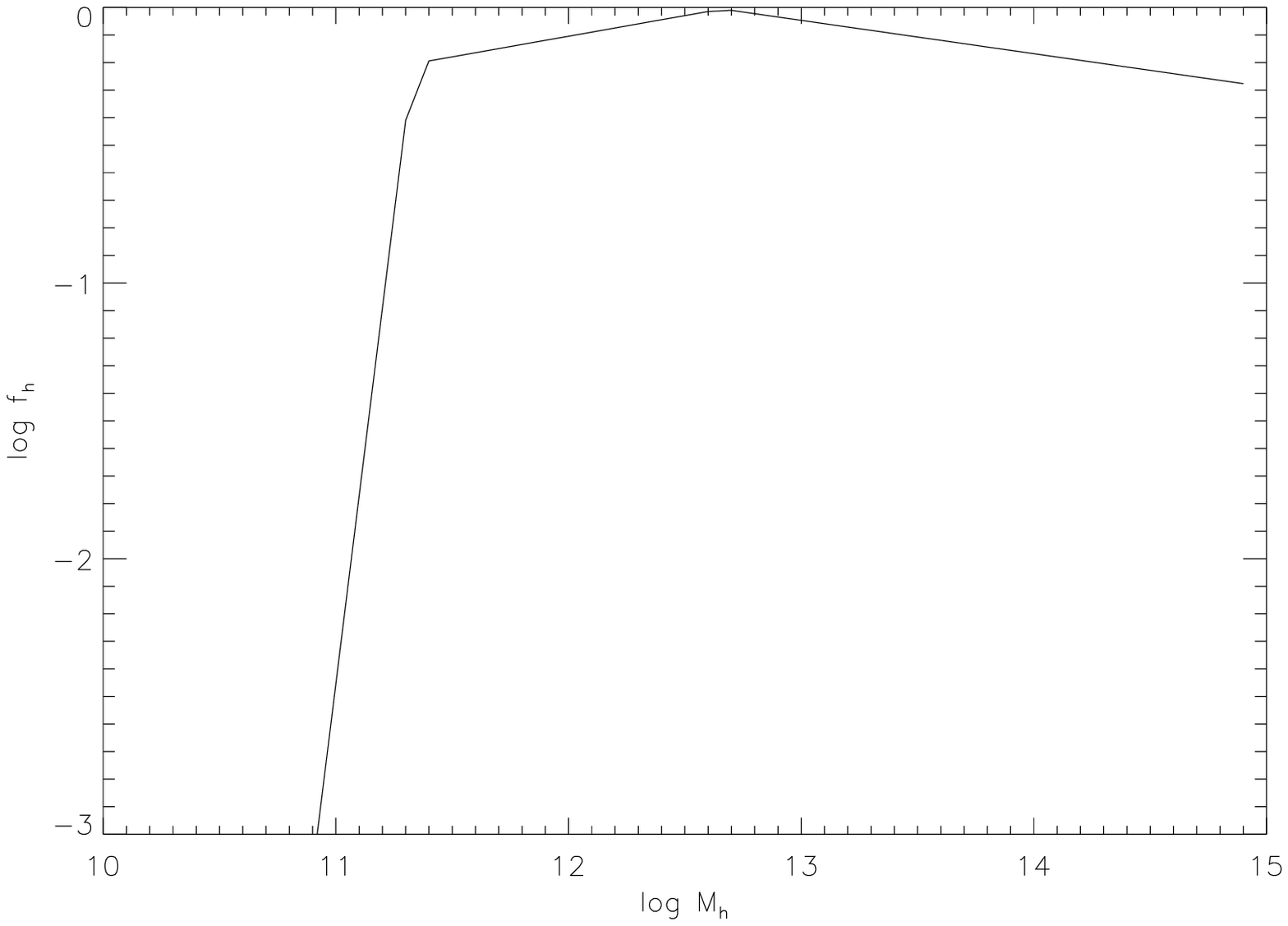}
}
\subfigure{
 \label{massFunc}
 \includegraphics[width=84mm]{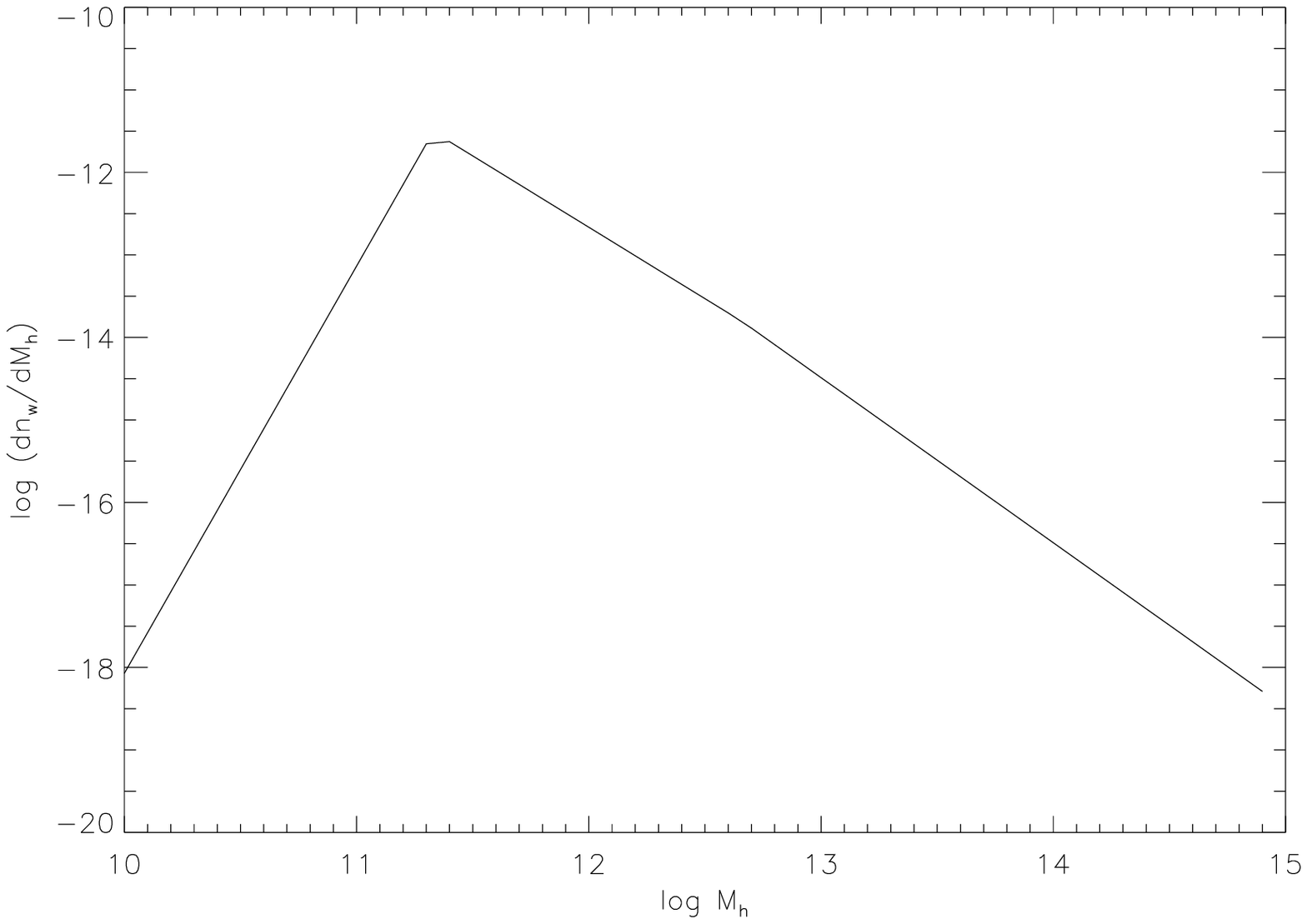}
}
\caption{{\bf Top} The probability that a dark matter halo of mass $M_h$ will show \mgii\ absorption, $f_h$, plotted against halo mass.  Note that at low masses, the probability is very low.  Much of the \mgii\ absorbing gas is found in dark matter halos with masses $>10^{11}M_{\odot}$. The form of $f_h$ was found by linearly interpolating between 4 values determined by the best fitting procedure in TC08.  {\bf Bottom} The weighted halo mass function plotted against halo mass.  The low value of $f_h$ at low masses forces $dn_w/dM_h$ to be relatively low at low masses, and the power law form of the halo mass function forces $dn_w/dM_h$ to be relatively low at high masses.}
\label{groupmassFunc}
\end{figure}

Instead of parameterizing the mass dependence of $f_h$, TC08 specified the values of $f_h$ at four different masses and interpolated between them.  The four masses used were $\log M_i/M_{\odot}=$ 10.0, 11.33, 12.66, and 14.0.  The four corresponding values of $f_h$ are denoted $f_1, f_2, f_3$ and $f_4$.  In TC08, the best fit values for $f_h$ at each $\log M_i$ were determined; their values were: $\log f_1=-9.27$, $\log f_2=-0.205$, $\log f_3=-0.006$, and $\log f_4=-0.168$. We adopt the same parameters and interpolate linearly between them; the plot in Figure \ref{fhprofile} shows the dependence of $f_h$ on mass (both in the log).

For the sake of simplicity, we approximate the halo mass function as a power law ($dn/dM_h=M_h^{-\alpha}$) that we sample over the range $10.75 < \log M_h/M_{\odot} < 14.75$.  In \nocite{WAHL06}{Warren} {et~al.} (2006), the parameterization is actually a power law combined with an exponential decay at high masses ($>10^{14.5} h^{-1} M_{\odot}$), however, our sampling of masses in this range is negligible and so we consider our approximation to be sufficiently accurate.  From the power law portion of the \nocite{WAHL06}{Warren} {et~al.} (2006) halo mass function, we measure $\alpha=1.88$.  Therefore, our weighted halo mass function is defined as
\begin{equation}
\frac{dn_w}{dM}= \frac{f_h M_h^{-\alpha}}{\int_{10.75}^{14.75}f_h M_h^{-\alpha} dM_h }
\label{massFunceqn}
\end{equation}
%
%
%
such that we have a normalized weighted halo mass function with sampling range $10.75 < \log M_h/M_{\odot}  < 14.75$.  A plot of $dn_w/dM_h$ is located in Figure \ref{massFunc}, where $n_w$ represents the weighted number of absorbers.

As noted in TC08, the weighted mass function peaks at $\log M_h/M_{\odot} \sim 12$, and nearly all halos of $11.5 < \log M_h/M_{\odot} < 12.5$ are expected to host \mgii\ absorption.  At lower masses, the halo mass function has a relatively high normalization but the low value of $f_h$ makes the probability of finding an \mgii\ absorber in the dark matter halo low, and thus we have a small chance of observing them.  At higher masses, $f_h$ is relatively high but the number of halos is much lower, making the probability of observing a \mgii\ absorber at such halo masses very unlikely.

\begin{figure}
\includegraphics[width=150mm]{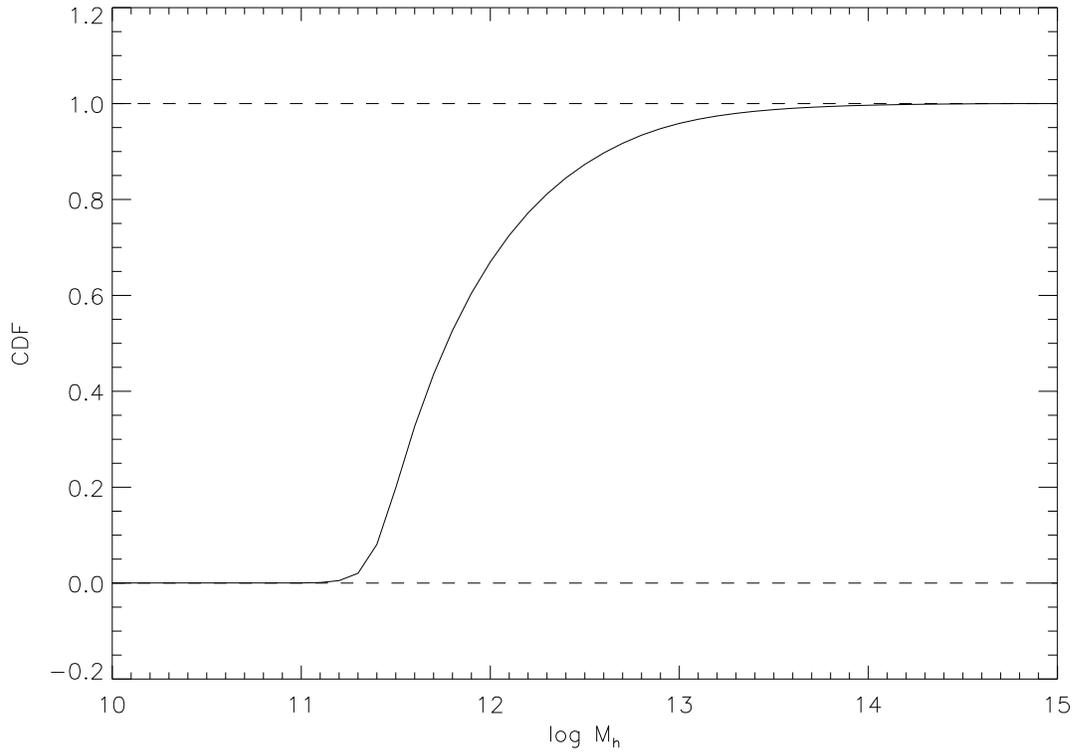}
\caption{The Cumulative Distribution Function found by integrating over $dn_w/dM_h$ for all values of mass.  Masses in the range $11 < \log M_h/M_{\odot} < 13$ are much more likely to exhibit \mgii\ absorption compared to all other masses (high or low).}
\label{CDF}
\end{figure}

The Cumulative Distribution Function (CDF) of the weighted halo mass function is plotted in Figure \ref{CDF}.  This plot shows that the masses we are most likely to be studying are approximately $11 < \log M_h/M_{\odot} < 13 $.  All other masses have a low probability of exhibiting \mgii\ absorption, although we still sample over the range $10.75 < \log M_h/M_{\odot}4 < 14.75$.  The CDF allows us to randomly select dark matter halo masses to properly represent the statistical distribution of dark matter halos in the universe that will harbour \mgii\ absorbing gas.  This model assumes no evolution of the halo mass distribution of absorbers with redshift, which has been suggested by observational evidence in \nocite{LWP11}{Lundgren} {et~al.} (2011).

\subsubsection{Simulations for Pairs}\label{simforpairs}

The location (i.e., impact parameter) and orientation of an observed quasar asterism in reference to its absorbing gaseous halo is unknown.  Here we describe how we generate a random distribution of simulated data within the constraints of the TC08 model that incorporates our ignorance of the location of the \mgii-absorbing galaxy.  We also describe how we test this distribution of simulated data against the observed data in Table \ref{pairstable}.  It is important to note that when simulating EWs we search for matches to both LOSA and LOSB in the observed quasar pair.  The only stipulation is that the observed EW in the LOS must be measured to better than $3\sigma$ (i.e., we do not match to upper-limits).

Each run of simulations begins by choosing and fixing the characteristic length and covering fraction combination to be used during the run.  These are chosen from within the ranges given at the end of \S\ \ref{TC08framework}.  We then generate a halo mass $M_h$ by selecting at random from the weighted halo mass function described in the previous section.  The mass is used to calculate $R_g$ using equation (\ref{Rg_new}), $R_{sh}$ using equation (\ref{shockradius}), and $G_0$ using equation (\ref{G0}).  An impact parameter $s_a$ is then randomly generated within the cloud of radius $R_g$, defining a location $a$ (all parameters defined at this location will have the subscript $a$).  Using these model parameters, impact parameter $s_a$, and equations (\ref{W1}-\ref{W3}) we generate $EW(s_a|M_h)$, the EW due to \mgii\ absorption at impact parameter $s_a$ given the halo mass $M_h$.  

In \nocite{CT08}{Chen} \& {Tinker} (2008), the authors found the model used in this study had a cosmic scatter ($\sigma_{cs}$) of 0.25 dex.  To incorporate $\sigma_{cs}$, we must scatter the simulated $EW(s_a|M_h)$ {\it individually} for each absorption system in our quasar sample.  For each line of sight in a pair that is measured to better than $3\sigma$ the scattered EW is
\begin{equation}
\log EW_a = \log EW(s_a|M_h) + \sigma_a (1-e^{-d_{x}/\ell_c}) + \sigma_x e^{-d_{x}/\ell_c}
\label{WA}
\end{equation}
where $\sigma_a$ is the cosmic scatter at the location $a$, $x$ denotes the point on the sky midway between the lines of sight of the observed pair, $\sigma_x$ is the cosmic scatter at $x$, and $d_{x}$ is the proper separation between $a$ and $x$ at the absorber redshift.
The values $\sigma_a$ and $\sigma_x$ are chosen at random from a Gaussian with $\mu=0$ and $\sigma=0.25$ dex.  The characteristic length $\ell_c$ is one of the variable parameters in the model.  Note that when $d_{x}\gg\ell_c$, the $\sigma_x$ term is negligible and the cosmic scatter is $\sigma_a$ at $a$.  If $d_{x}\ll \ell_c$ then $\sigma_x$ dominates the cosmic scatter at $a$, while the $\sigma_a$ term is negligible.  This relationship is used because we assume there is some scale on which cosmic scatter should not apply, and the two lines of sight will be, effectively, identical.  However, if the separation of the pair is much larger than the characteristic length, then cosmic scatter may be applied to each LOS calculation individually.

Note that the value $d_x$ in equation (\ref{WA}), is unique to each absorption system, and thus the value of $EW_a$ is calculated individually for each entry in Table \ref{pairstable} before checking to see if it is a match.

Using $EW(s_a|M_h)$ and equation (\ref{WA}), we search the different lines of sight in the pair sample for an EW that matches the scattered $EW_a$ (note that we can match {\it either} LOSA or LOSB to this value).  To be considered a match the simulated value must be within $\pm\sqrt{3}\sigma$ of the given observed EW (where $\sigma$ is the observational uncertainty on the EW; see Table \ref{pairstable}).\footnote{This range is used because a box function of width $2\sigma\sqrt{3}$ has the same variance as a Gaussian function with variance $\sigma^2$.}

In cases where the EW in both lines of sight at a given redshift in an observed pair are measured at $>3\sigma$, we check for matches with both the first line of sight (LOSA) and the second line of sight (LOSB) in the pair individually (see Table \ref{pairstable} for which LOS is 'A').  In cases where only the EW in LOSA is measured to a statistically significant amount, and LOSB has been quoted as an upper limit, we search for matches to LOSA only (and vice versa).  The cases for triplets and quads will be described later (see section \ref{triplesquads}).  It is entirely possible that a randomly generated $EW(s_a|M_h)$ does not scatter to match the EW at either LOSA or LOSB in a given absorption system.  

When a match to an observed LOS is made, we use the TC08 model to compute a second EW, located at $s_b$, which is a distance $d$ away from the matched LOS located at $s_a$ (The distance $d$ is the separation listed in column 3 of Table \ref{pairstable}).  This second EW is generated by
\begin{equation}
\log EW_b = \log EW(s_b|M_h) +\sigma_b (1-e^{-d_{x}/\ell_c}) + \sigma_x e^{-d_{x}/\ell_c}.
\label{WB}
\end{equation}
%
%
The impact parameter $s_b$ is the location of the second line of sight in the quasar pair, and is found by placing LOSB at a separation of $d$ from LOSA and randomly rotating it in the plane of the sky an angle of $0<\phi<2\pi$.  The value $EW(s_b|M_h)$ is then determined using equations (\ref{W1}-\ref{W3}).  The value $\sigma_b$ is found in the same way as $\sigma_a$ and $\sigma_x$.\footnote{The value of $\sigma_x$ is the same for both $EW_a$ and $EW_b$.}  The resulting $EW_b$ value is the TC08 predicted EW for a LOSB in a quasar pair, given a match to LOSA ($EW_a$) and the proper separation of the pair.  Again, note that we also allow for the reverse: matching to LOSB and predicting LOSA.
In the limit of $d_x=0$, $s_a=s_b$ ($a$ and $b$ represent the same point) so that $EW(s_a|M_h)=EW(s_b|M_h)$ and $\log EW_a=\log EW(s_a|M_h)+\sigma_x=\log EW_b$, as expected.  Application of the covering fraction is discussed in \S\ \ref{cov_frac}.

For each line of sight in the 18 absorption systems in the quasar pair sample that was {\it not} measured as an upper limit, we tabulated 900 simulated $(EW_a,EW_b)$ pairs.  Each set of 900 contains simulated EWs that have the same corresponding proper separation as the observed absorption system, but random halo masses, projected impact parameters, and orientations on the plane of the sky.  Once each LOS in the pairs sample has tabulated 900 pairs, we re-run the simulations for a new combination of characteristic length and covering fraction.

\subsubsection{Simulations for Triples and Quads}\label{triplesquads}
Applying the model to a quasar triple or quad requires a slightly different approach, in that there are now three or four lines of sight to consider when predicting EWs.

The equation for $EW_a$ is conceptually the same as for pairs.  A random location in the absorbing halo is chosen.  We define its impact parameter to be $s_a$, and we have:
\begin{equation}
\log EW_a = \log EW(s_a|M_h) + \sigma_a (1-e^{-d_{ax}/\ell_c}) + \sigma_x e^{-d_{ax}/\ell_c}
\label{quadA}
\end{equation}
where $\sigma_a$ is the cosmic scatter at the location given by $s_a$, $d_{ax}$ is the proper distance from the LOS to the projected centre $x$ of the quasar asterism (see Table \ref{quadEWtable}), and $\sigma_x$ is the cosmic scatter at $x$.  The values $\sigma_a$ and $\sigma_x$ are chosen at random from a Gaussian with $\mu=0$ and $\sigma=0.25$ dex.

In each triple/quad there are multiple lines of sight per absorption system.  We check each LOS for a match to the random EW, with the same stipulations for a match as used by the pairs simulation.  When a match is made to one of the observed lines of sight in the asterism, we place a scaled, rotated copy of that asterism in the gas cloud, with the LOS to which a match has been made located at impact parameter $s_a$.  The scaling takes into account that the proper distances between the lines of sight are dependent on the redshift of the absorption feature. The asterism copy is rotated by a randomly chosen angle $\phi$ between 0 and 2$\pi$.  To calculate the $EW_n$ values of the other lines of sight in the asterism, we use
\begin{equation}
\log EW_n = \log EW(s_n|M) +\sigma_n (1-e^{-d_{nx}/\ell_c}) + \sigma_x e^{-d_{nx}/\ell_c}
\label{tripquadW}
\end{equation}
where $n=b$, $c$, or $d$, $d_{nx}$ is the equivalent of $d_{ax}$ for each sight-line and $\sigma_n$ is the equivalent of $\sigma_a$ for each sight-line.  The values $EW(s_n|M_h)$ are determined by equations (\ref{W1}-\ref{W3}).  The fourth EW ($EW_d$) is only generated if the object is a quad.  This method generates three or four EWs in the same geometry as an observed object, at a random point inside a gaseous halo described by one randomly selected mass.  For each LOS with an EW measured at $>3\sigma$ in each object, we tabulate $900$ sets of simulated EWs.

\begin{figure}
\includegraphics[width=150mm]{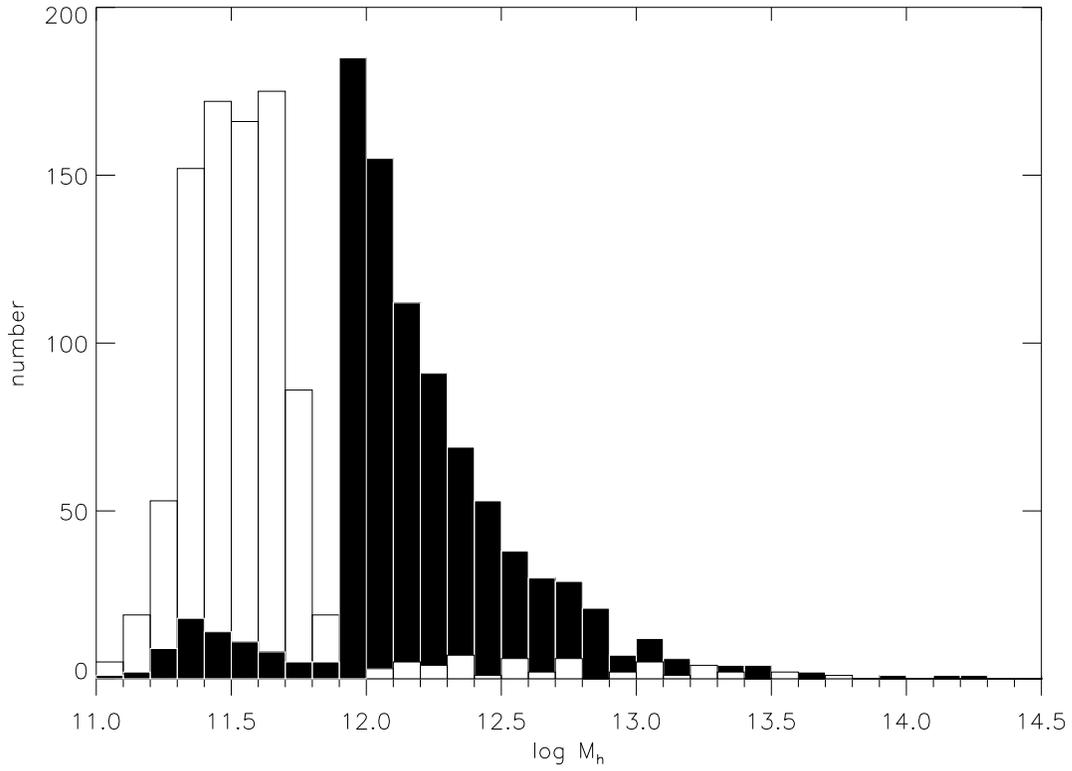}
\caption{The distribution of masses generated by our simulations that resulted in a match to a line of sight.  The white histogram is for a strong absorber with $EW = 0.80 \pm 0.2$~\AA.  The black histogram is for a weak absorber with $EW = 0.05 \pm 0.01$~\AA.  Both absorbers are from APM08279$+$5255 (Ellison et al. 2004).  Each histogram totals 900 different masses.
}
\label{weakVSstrong}
\end{figure}

As a sanity check, in Figure \ref{weakVSstrong} we have plotted the distribution of masses generated by the simulations that ended in a match to the EW in a line of sight from the data set.  We plotted the distribution for two different lines of sight, a weak absorber and a strong absorber.\footnote{The dividing line between 'weak' and 'strong' has been defined in the literature as $EW=0.3$\AA.}  Both absorbers were from the APM08279+5255 simulations.

The two mass distributions are clearly distinct, but keeping Figure \ref{EWprofile} in mind, they make sense.  The less massive halos have TC08 modeled EW profiles that are more able to achieve larger EW values.  In randomly generating masses and looking for matches, it makes sense that the distribution for a strong absorber will include more massive halos.

\subsubsection{Covering Fraction}\label{cov_frac}
We have not yet accounted for the covering fraction.  Effectively, the above description is one of simulating line of sight EWs with 100\%\ covering fraction.  As discussed in section \ref{intro} the measurements of $f_c$ have a wide range of values.  We test a range of covering fractions from 0.01 to 1.0, which encompass the results of \nocite{KCS08}{Kacprzak} {et~al.} (2008) ($f_c=0.5$ for $0.02 {\rm~ \AA}< EW < 3.0$~\AA) and \nocite{BC09}{Barton} \& {Cooke} (2009) ($f_c\le0.4$ for weighted mean rest-frame $EW=1.29$~\AA) as well as the covering fraction measured from this model by \nocite{CT08}{Chen} \& {Tinker} (2008), $f_c=0.85$ for weighted mean rest-frame $EW=0.34$~\AA.  Note that the covering fraction measured in these studies depends on the EW range of the data set, with smaller $f_c$ at higher EW.

Each simulation run has a fixed characteristic length and covering fraction.  When a match to an observed sight-line is made by scattering $EW(s_a|M_h)$, each of the other sight-lines in the quasar asterism would then have a $f_c$ chance of actually absorbing in \mgii.  (We do not apply the covering fraction to sight-line A because we are measuring the ability of the TC08 model to predict the EW in other sight-lines {\it given} the matched EW at LOSA.)

For pairs, to account for the covering fraction in the LOS located at position $b$, we randomly generate a number between 0 and 1 for each of the 900 $(EW_a,EW_b)$ simulated pairs.  If the random number is above the covering fraction value, we reset $EW_b$ to the value $EW_b^{\prime}$ given by
\begin{equation}
EW_b^{\prime} = \left[EW_a + \sigma_a \right]e^{-d/\ell_c}.
\label{WBprime}
\end{equation}
where $EW_a$ is the matched EW, $\sigma_a$ is the cosmic scatter at $a$, and $d=2d_x$ is the separation between LOSA and LOSB.  
In the case of triples and quads, we also recalculate the EW at each of the lines of sight predicted by the TC08 model using a random number approach, however, we do this relative to the EW at the centre of the asterism:
\begin{equation}
EW_n^{\prime} = \left[EW(s_x|M_h) + \sigma_x \right]e^{-d_{nx}/\ell_c}
\label{Wiprime}
\end{equation}
where $n=b,c$ or $d$, $EW(s_x|M_h)$ is the EW at $x$, $\sigma_x$ is the cosmic scatter at $x$, and $d_{nx}$ is the distance from LOSn to $x$.

As a result, for all quasar asterisms in our tabulated data from the literature we have created $900$ unique and independent $(EW_a,EW_n)$ sets at each value of $\ell_c$ and $f_c$ with the value $EW_a$ matched to one of the lines of sight of the parent quasar asterism.  Each of the 900 simulated sets was generated for a randomly chosen dark matter halo mass, projected impact parameter, orientation, and probability $f_c$ of absorbing in \mgii at a given location.


\section{Analysis: Comparing the Simulations with the Observations}\label{analysis}
In this section we describe how we analyze the simulations.  We use the Kolmogorov-Smirnov (KS) test (in \S\ 4.3) to compare the simulations to what is expected if the TC08 model exactly predicts the observed distribution of equivalent widths.  To that end, we take into account the differences in relative size of the gaseous halos used to generate the simulated EWs.  We use the cumulative distribution function of the fractional probabilities for each paired sight-line (\S\ 4.2) as a measurement of the ability of the TC08 model to predict the observed EWs.

The original TC08 model was constrained using a sample ranging in redshift over $0.2067 < z < 0.892$.  We provide the KS test for both our entire sample, and a redshift limited sample in the same range originally used in the TC08 model.

For illustrative purposes, below we denote an observed EW as $EW_o\pm\sigma_o$ and its corresponding 900 simulated EWs as $EW^{sim}_j$.

\subsection{Relative Size Considerations}

Because gaseous halos of different masses and thus sizes were used to create the distribution of $EW^{sim}_j$ for each $EW_o\pm\sigma_o$, we must take into account the relative probability each individual halo has of producing absorption that matches an observation (call it $P^\prime_j$).  Each halo mass has a unique gaseous halo size as determined by $R_g$, a unique equivalent width profile (see Figure \ref{EWprofile}), and a cosmic scatter of 0.25 dex in the EW at any given position.  Therefore, each impact parameter in each gaseous halo has a different probability of producing an absorption feature that matches an absorption feature from the observed data set.  

For any one impact parameter $s$ relative to the centre of the $j$th gaseous halo of mass $M_j$ and radius $R_j$, we weight that impact parameter following
\begin{equation}
2\pi s ~\cdot P_{abs}(EW_o\pm\sigma_o|s)
\label{sweight}
\end{equation}
where $P_{abs}(EW_o\pm\sigma_o|s)$ is the probability of generating an equivalent width match to an observation of $EW_o\pm\sigma_o$, given $s$ and the equivalent width profile of the gaseous halo.

\begin{figure}
\includegraphics[width=150mm]{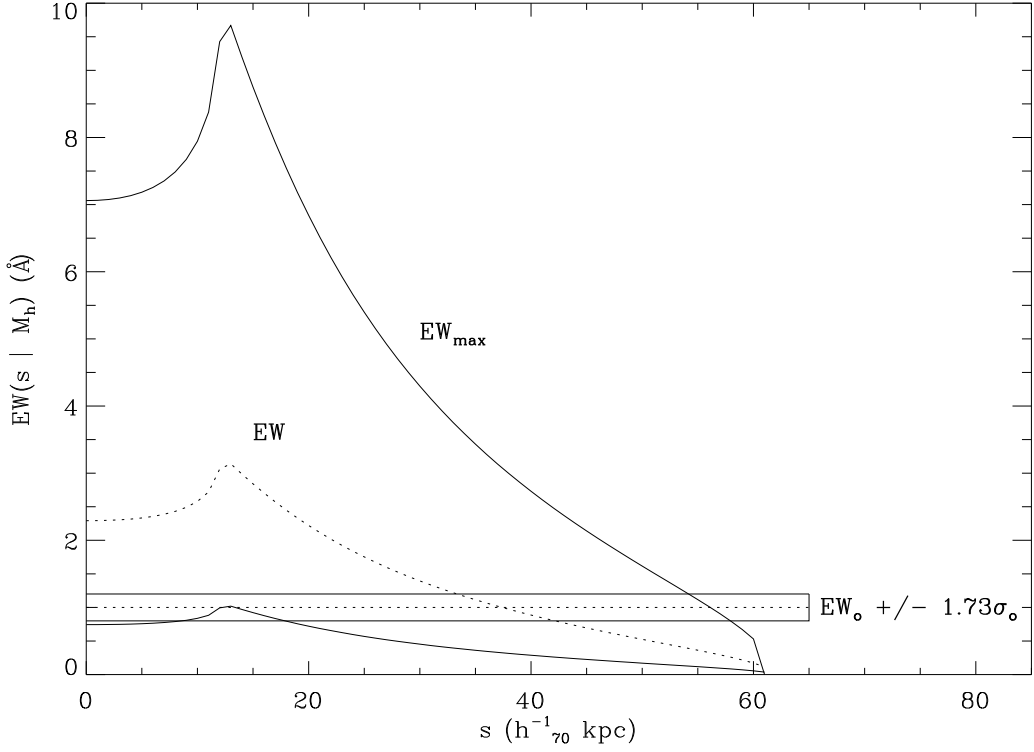}
\caption{The dotted line labeled '$EW$' represents the unscattered TC08 model for $M=10^{11.2} M_{\odot}$.  The solid line labeled '$EW_{\rm max}$' represents the 1$\sigma$ cosmic scatter above the TC08 model; the cosmic scatter below the TC08 model is also whosn as a solid line.  An observed \mgii\ absorber is labeled $EW_o \pm 1.73\sigma$, where $\sqrt{3}\sim1.73$.  The region where the the observed \mgii\ system and the scattered TC08 model overlap represents the relative probability that a gaseous halo defined by the above mass will generate a simulated equivalent width that matches the observed value.}
\label{Pabsfigure}
\end{figure}

To determine $P_{abs}(EW_o\pm\sigma_o|s)$, we first need $P_{abs}(s)$, the probability distribution of EWs at $s$ for a given $M_j$ in the TC08 model.  The exact form of $P_{abs}(s)$ is
\begin{equation}
P_{abs}(s) = D\exp\left( \frac{-[\log EW-\log EW_r(s|M_j)]^2}{2\sigma_{cs}^2} \right) 
\label{Pabs}
\end{equation}
where $EW_r(s|M_j)$ is the simulated equivalent width at the given impact parameter (see equations \ref{W1}-\ref{W3}), $\sigma_{cs}=0.25$ and the constant $D$ is a normalization factor. 
We approximate the Gaussian in equation (\ref{Pabs}) as a uniform probability distribution in $\log~r$ between $\log EW_{min}=\log EW_r(s|M)-\sigma_{cs}\sqrt{3}$ and $\log EW_{max}=\log EW_r(s|M)+\sigma_{cs}\sqrt{3}$.
%
%
To determine the relative probability that the individual impact parameter $s$ will produce an absorption feature that matches an observation $EW_o\pm\sigma_o$ we measure the area between $EW_{min}$ and $EW_{max}$ that overlaps the allowed range around the observation.  For each individual impact parameter, the area where the bounds of the observed $EW_o\pm\sigma_o$ and the bounds of the TC08 model overlap represents the relative probability of producing an absorption feature that would be considered a match:
\begin{equation}
P_{abs}(EW_o\pm\sigma_o|s) = \int^b_a{\frac{d\log EW}{\log EW_{max} - \log EW_{min}}}
\label{PabsapproxWi}
\end{equation}
\begin{equation*}
{\rm where} \;\; a = \max(\log EW_{min}, \log EW_o-\sigma_o\sqrt{3})
\end{equation*}
\begin{equation*}
{\rm and} \;\;b=\min(\log EW_{max}, \log EW_o+\sigma_o\sqrt{3}).
\end{equation*}
In Figure \ref{Pabsfigure} we show an example of this calculation.  An observed equivalent width, labeled $EW_o$, has some range over which we consider it to be matched, determined by the observed uncertainty $\pm\sigma\sqrt{3}$.  In the case of Figure \ref{Pabsfigure}, the equivalent width profile $EW$ was created using a halo mass of $\log M_h/M_{\odot} = 11.5$.

As Figure \ref{Pabsfigure} shows, it is possible to match $EW_o$ over a range of impact parameters, each of which contributes to the probability of matching the observation in a halo of size $R_j$; we thus sum over all impact parameters within the $j$th gaseous halo to find the relative probability of obtaining a match to the observed value $EW_o$ in that halo.  
%
%
We then divide by the sum of the relative probabilities of all $N$ simulated matches to find $P^\prime_j$, which is the relative probability of matching the observed $EW_o$ in a halo with radius $R_j$:
\begin{equation}
P^\prime_j= ]frac{\displaystyle\sum_{s<R_j}2\pi s~\delta s~\cdot P_{abs}(EW_o\pm\sigma_o|s)}{\displaystyle\sum_{k=1}^N \left(\displaystyle\sum_{s<R_k}2\pi s~\delta s\cdot P_{abs}(EW_o\pm\sigma_o|s) \right ) }
\label{sweightsumtotal}
\end{equation}
where $N$ is the number of simulated matches.  

Each of the 900 $EW^{sim}_j$ are no longer weighted equally at $1/900$, but now weighted as $P^\prime_j/900$.  As a simple example, if $P_{abs}(EW_o\pm\sigma_o|s)={\rm fixed}$ independent of $s$ or $M_j$ then the weight of each halo is just $\propto R_j^2$, and the relative weight of each halo is $P^\prime_j=R_j^2/(\sum_j R_j^2)$.  

\subsection{Fractional Probability Curves}

For each choice of $\ell_c$ and  $f_c$, we measure how well the simulated EWs match the observed measurements by looking at the cumulative distribution function (CDF) of the fractional probabilities for each paired sight-line.  Conceptually, the fractional probability is the fraction of the 900 simulated EWs which are equal to or less than\footnote{The fraction equal to or greater than the actual measured EW is an equally good statistic.} the actual measured EW of LOSB in that paired sight-line.  If the model being tested matches the observations well, then the fractional probabilities generated by that model for our sample of paired sight-lines should be uniformly distributed between 0 and 1.

However, some of the $EW_o$ values are measured as upper limits and the remaining have uncertainty values.  Measuring simply the number of values below $EW_o$ does not include the observational uncertainties associated with it.  To account for this, we compared the weighted cumulative distribution of the $EW^{sim}_j$ (weighted by $P_j^{\prime}$)to the cumulative distribution of a Gaussian with $\mu=EW_o$ and $\sigma=\sigma_i$; i.e., an error function erf($\mu,\sigma$).\footnote{If the observed EW is actually an upper limit, we set $\mu=0$ and $\sigma=\sigma_o$, where $\sigma_o$ is equal to one third the value of the upper limit quoted in Table \ref{pairstable}.}  The erf($\mu=EW_o$, $\sigma=\sigma_o$) values plotted as a function of the weighted cumulative distribution of the $EW^{sim}_j$ provide a fractional probability curve for each paired sight-line.  That curve combines the fractional probability at each equivalent width with the relative probability of each equivalent width being the true value given the observed $EW_o$ and $\sigma_o$.  Thus, instead of generating the CDF of fractional probabilities for all sight-lines by summing step functions corresponding to a single fractional probability for each paired sight-line, we generate it by summing the fractional probability curves for each paired sight-line.

\subsection{The Kolmogorov-Smirnov Test}

We can use the Kolmogorov-Smirnov \nocite{numRec07}(Press {et~al.} 2007) test to see how well the simulated TC08 model equivalent widths match the observations.

The Kolmogorov-Smirnov (KS) test is a way of measuring the difference between two CDFs.
The amount by which the CDF $S_N(x)$ of some given data set differs from a known distribution denoted $P(x)$ is measured by finding the maximum value of the absolute difference between the two CDFs.  This is denoted $D$ and defined as
\begin{equation}
D = \max \left | S_N(x) - P(x) \right | .
\label{KSD}
\end{equation}  
To avoid decreased sensitivity near $P(x)=0$ or $1$, we use the variant of the KS test known as the Kuiper statistic \nocite{numRec07}(Press {et~al.} 2007).  The Kuiper statistic $V$ measures the maximum value of the difference in the CDFs above and below $P(x)$:
\begin{equation}
V= \max \left [ S_N(x) - P(x) \right ] + \max \left [ P(x) - S_N(x) \right ].
\label{KTD}
\end{equation}

If the simulated data generated by the TC08 model above matches well the observed data points from the literature, we would expect the sum of the fractional probability curves of the individual observed EWs to create a cumulative distribution function that is similar to $P(x)=x$.  We can measure the deviation from $P(x)=x$ using the Kuiper statistic.  The model parameters that yield the smallest value of $V$ therefore represent the parameters that best match the observed values of EW.

We seek to determine how well each combination of the model parameters $\ell_c$ and $f_c$ reproduces the observed data.  For this analysis, we separated the triples and quads into sets of pairs.  A triple quasar can be considered a set of 6 individual quasar pairs.  A quad can be considered a set of 12 individual quasar pairs.\footnote{This approach is conservative in the sense that we do not assume that matching LOSA and predicting LOSB yields the same results as matching LOSB and predicting LOSA.}  In doing so we do not lose information regarding the two dimensionality of the asterisms, as the EWs of each line of sight were generated keeping the form of the asterism in the gaseous cloud consistent.  We only break the asterisms apart in order to calculate individual fractional probability curves.


In total, there are 25 fractional probability curves from the pair sample, 48 from the triples sample, and 39 fractional probability curves from the quads sample, with a total set of 112 individual fractional probability curves per combination of $(\ell_c, f_c)$.  

\begin{figure}
\includegraphics[width=150mm]{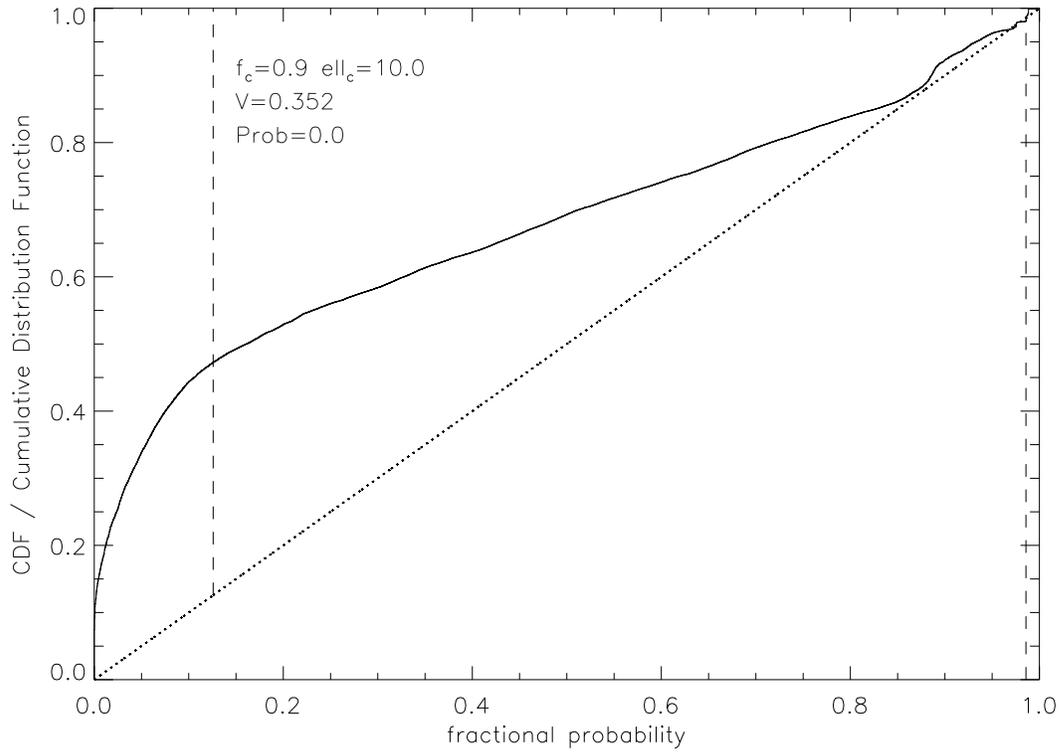}
\caption{An example of a bad fit to the null hypothesis.  The fractional probability curve is for $\ell_c=10.0~h^{-1}_{70}$~kpc and $f_c=0.9$. The p-value probability is effectively 0.0 indicating that, with these model parameters, the TC08 model cannot reproduce the observations.}
\label{exampleKStest_BAD}
\end{figure}

\begin{figure}
\includegraphics[width=150mm]{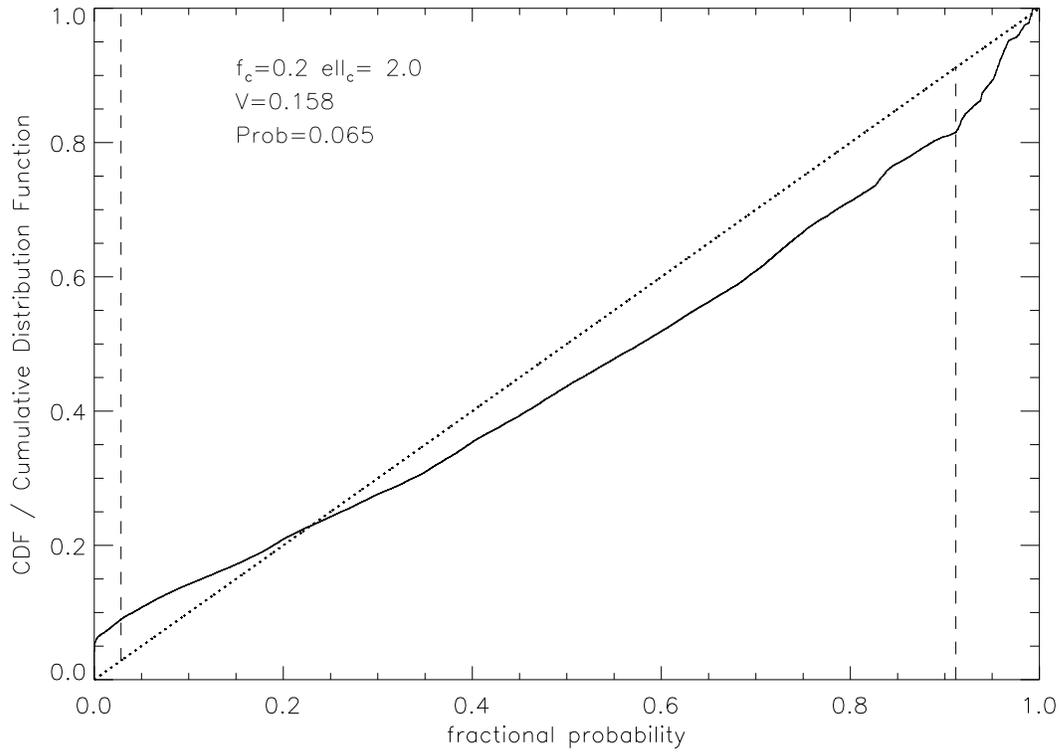}
\caption{The fractional probability curve for $\ell_c=2.0~h^{-1}_{70}$~kpc and $f_c=0.2$.  This is a much better fit than the model parameters in Figure \ref{exampleKStest_BAD}, and the p-value probability of $\sim0.13$ indicates that we cannot rule out these parameters to better than 3$\sigma$.}
\label{exampleKStest_MID}
\end{figure}

\begin{figure}
\includegraphics[width=150mm]{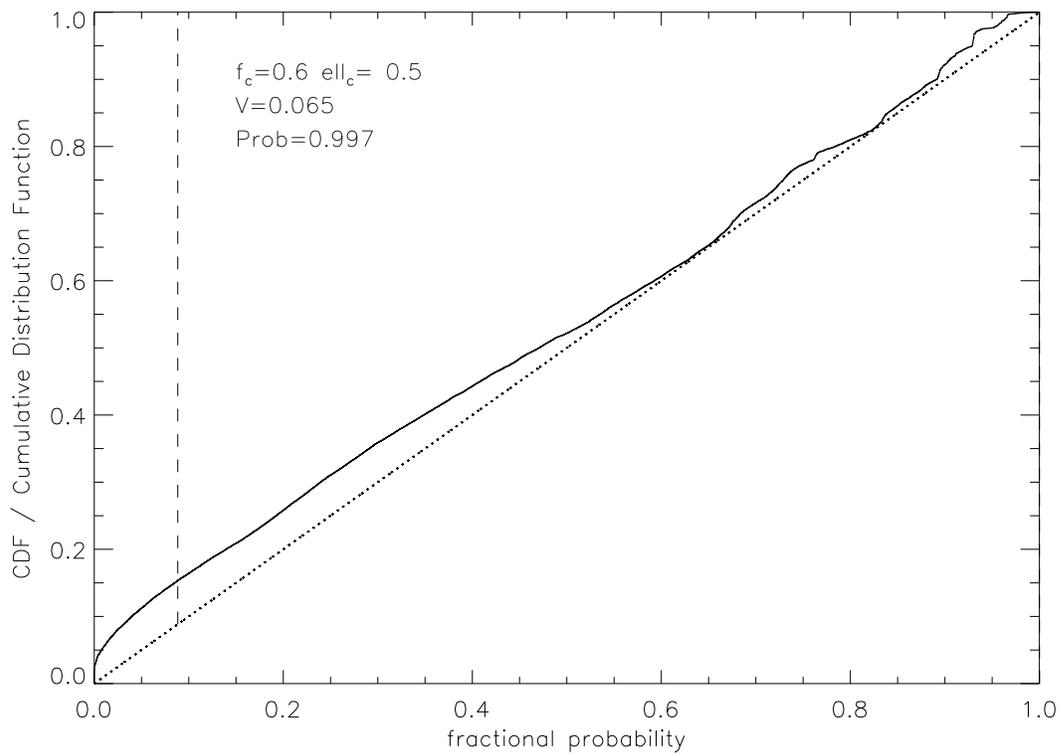}
\caption{The fractional probability curve for the best-fit characteristic length $\ell_c=0.5~h^{-1}_{70}$~kpc and covering fraction $f_c=0.6$.}
\label{exampleKStest_BEST}
\end{figure}

For Figures \ref{exampleKStest_BAD}, \ref{exampleKStest_MID}, and \ref{exampleKStest_BEST}, the dotted line represents the null hypothesis: if the model exactly matches the observed data set, we expect the fractional probabilities to create a curve of $P(x)=x$.  The solid line is the fractional probability curve from the simulated data.  The vertical dashed lines show the positions of the largest positive and negative deviations of the histogram from the $P(x)=x$ curve.  The $V$ statistic (see equation \ref{KTD}) and the p-value are both labeled in the top left corner of the graphs.  The p-value is the probability, given the data, of determining a value for $V$ greater than the one found, adopting the null hypothesis.  

\begin{figure}
\includegraphics[width=150mm]{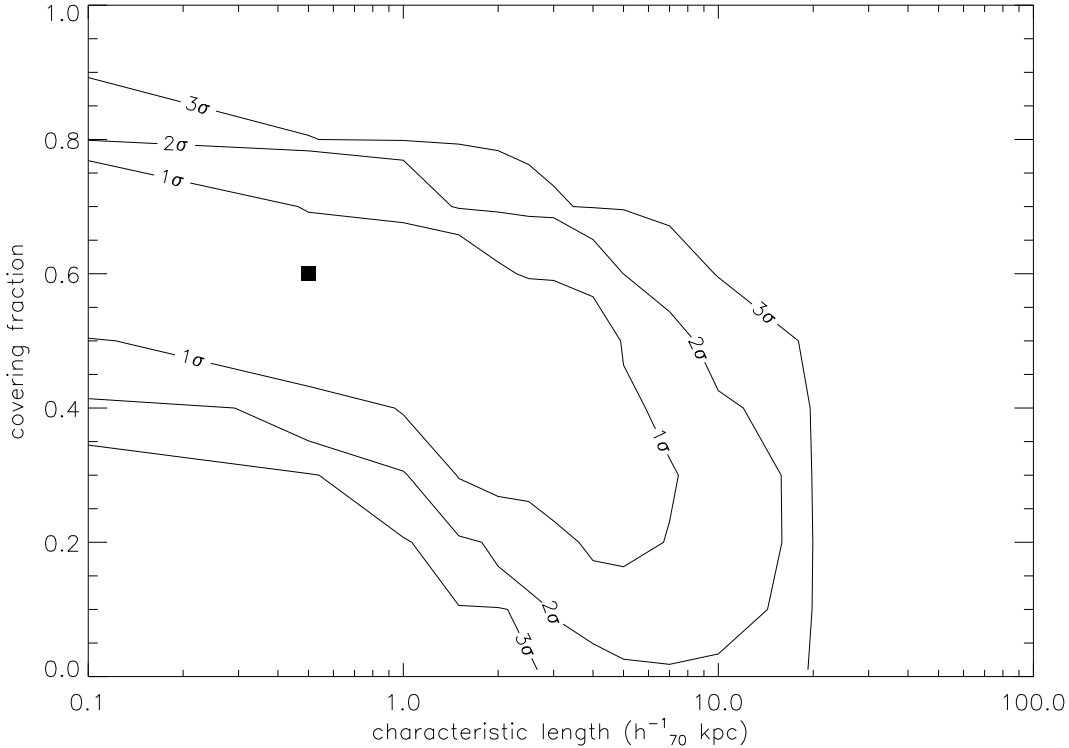}
\caption{Contour plot showing the results of applying the KS test with Kuiper variant to each combination of $(\ell_c,f_c)$.  The 1$\sigma$, 2$\sigma$, and 3$\sigma$ contours represent regions outside of which we can rule out the TC08 model to 68.3\%, 95.4\%, and 99.7\%\ confidence, respectively.  The black square denotes the best-fit characteristic length $\ell_c= 0.5~ h^{-1}_{70}$~kpc and covering fraction $f_c=0.6$.}
\label{contour}
\end{figure}

For each point in $(\ell_c, f_c)$ space we calculate the p-value.  A contour plot in $(\ell_c, f_c)$ space is shown in Fig. \ref{contour}.
We are able to reject the TC08 model in all parameter space outside of the $3\sigma$ contour.  The $2\sigma$ and $1\sigma$ lines are also plotted.
The best fit to the observed data occurs at $\ell_c=0.5~h^{-1}_{70}$~kpc and $f_c=0.6$, measured for a mean rest-frame $EW=0.87$\AA. 

\subsection{A Redshift Limited Sample}

In TC08, the model was initially constrained using both a \mgii-LRG cross-correlation function built from a sample that ranged from $z=0.35 - 0.80$, and a \mgii\ absorption frequency distribution function compiled from two sources that spanned the range $z=0.2 - 2.2$ (see \S\ 3 of \nocite{TC08}{Tinker} \& {Chen} 2008).  The TC08 model was then tested using a sample of 13 galaxy/absorber pairs with an average absorber redshift of $\langle z_{abs} \rangle =0.3818$ ranging $z=0.2067 - 0.892$ \nocite{CT08}({Chen} \& {Tinker} 2008).  The study yielded a cosmic scatter of 0.25 dex, which we have adopted in this study.

The mean redshift of the \mgii\ absorbers in our sample was $\langle z_{abs} \rangle =1.099$ with a range of $z=0.043 - 2.066$.  Our sample probes to smaller redshifts than the TC08 model was initially constrained with, but the contribution of our sample to the $z<0.2$ regime is only 2 out of 38 absorbers or $\sim$5\%\ of the sample.  We were interested to see if using the subset of \mgii\ absorbers in our sample that were within the redshift range used to observe the 0.25 dex cosmic scatter would change the constraints on the model parameters we varied.

\begin{figure}
\includegraphics[width=150mm]{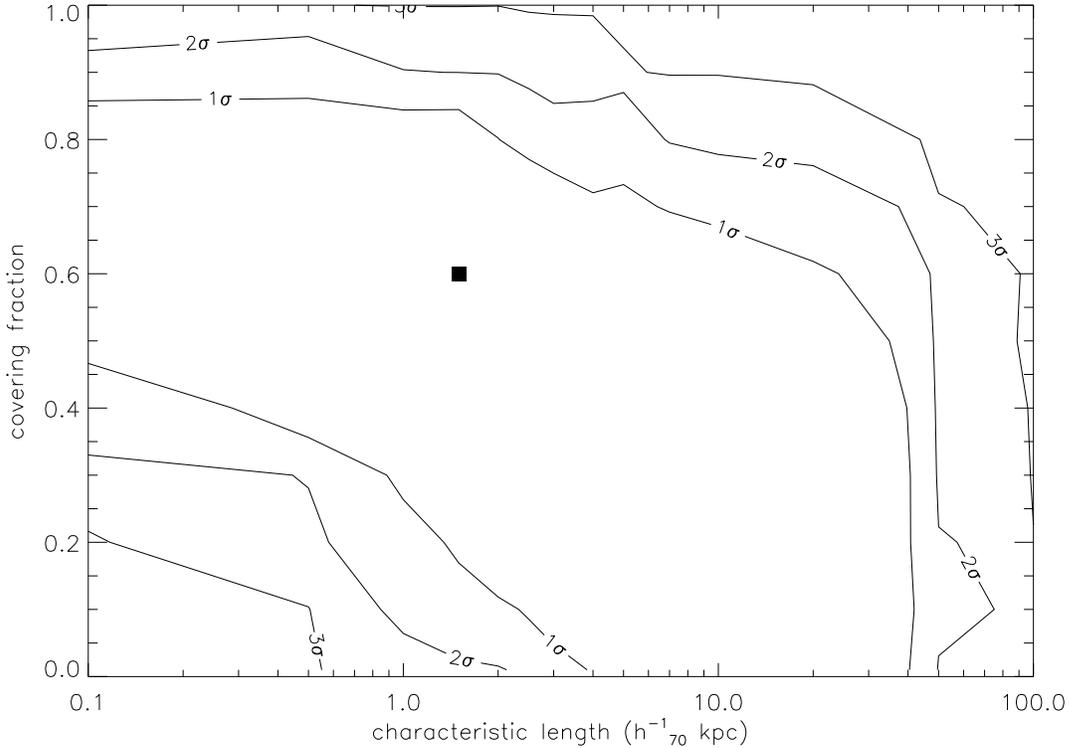}
\caption{Re-plotting the contours using only \mgii\ absorbers from our sample that lie within the redshift range $0.2067< z < 0.892$.  This range matches that used by Chen \&\ Tinker (2008) when testing the TC08 model.  The black square denotes the best-fit characteristic length $\ell_c=1.5 ~ h^{-1}_{70}$~kpc and covering fraction $f_c=0.6$.}
\label{contour_tc08}
\end{figure}

In Figure \ref{contour_tc08} we re-calculated the p-value probabilities and then re-plotted the contours using only simulations generated from observations whose \mgii\ absorption feature lies within $0.2067 < z < 0.892$.  By limiting the redshift range, we reduced our absorber sample from 38 to 12.  

The redshift limited sample is much less constraining than the entire sample (Figure \ref{contour}).  This is most likely due to the drastically reduced sample size.  The KS test with Kuiper variant is heavily dependent on the number of data points used, and the lack of constraint in Figure \ref{contour_tc08} is attributed to the lack of data points involved.

\section{Discussion}\label{discussion}
Our analysis has ruled out significant portions of the parameter space we tested.  All characteristic lengths larger than 20 $h^{-1}_{70}$ kpc are ruled out, regardless of covering fraction.  This result indicates that the underlying form of the TC08 model does not represent the distribution of the \mgii\ absorbing gas.  When the characteristic length is so large, it forces both sight-lines in a pair to be scattered by the same value $\sigma_x$.  This preserves the underlying form of the TC08 model (i.e., the 'unscattered' form), and is thus 'halo-to-halo' scatter; all sight-lines in one gaseous halo are scattered the same amount.  This is different from 'point-to-point' scatter, where all sight-lines in one gaseous halo are scattered by different amounts.  By ruling out large characteristic lengths, we have found that the TC08 model requires the cosmic scatter to be applied point-to-point, and thus the underlying analytical form of the model cannot fully reproduce the observed variation in \mgii\ absorption between sight-lines.


Given the above argument, it is not surprising that we have also ruled out covering fractions above 0.87 (for all characteristic lengths).  At covering fractions above 0.87, the probability that any sight-line will be set to zero is low, and thus we are forcing the use of the underlying analytical form of the TC08 model (see Figure \ref{EWprofile}).  Observations show that not all lines of sight below the estimated $R_g$ show \mgii\ absorption, and thus it is not surprising that high covering fractions are ruled out.

There is also a region at low characteristic lengths and low covering fractions that has been ruled out (see the bottom left of Figure \ref{contour}).  At covering fractions $f_c<0.3$, only the matched sight-line is guaranteed to have absorption; most other sight-lines only have residual absorption governed by the characteristic length, as set out by equations \ref{WBprime} and \ref{Wiprime}.  If the separation between sight-lines is much larger than the characteristic length of the simulation, most b, c, or d sight-lines will be set to zero.  In the literature this is clearly not the case, and thus we are able to rule out low covering fractions when the characteristic length is much smaller than the average separation of the data set.

As the characteristic length increases, the exponential factors of equations \ref{WBprime} and \ref{Wiprime} are no longer approximately zero.  As a result, even though most b, c, or d sight-lines are randomly chosen to be set to zero (at $f_c<0.3$), the large characteristic length forces many to be non-zero.  This effect is clearly visible in Figure \ref{contour}, where we can no longer rule out low covering fractions between $2.0 < \ell_c/ h^{-1}_{70}~ {\rm kpc} < 20$.

\section{Summary}\label{summary}

We have tested whether the model due to Tinker \&\ Chen of \mgii\ absorption from the gaseous halo around a galaxy can reproduce absorption in quasar pairs (both lensed and physical) and lensed triples and quads from the literature.  Specifically, we have determined what the acceptable values of the characteristic length and covering fraction. 

The TC08 model predicts the magnitude of \mgii\ absorption due to a gaseous halo due to a galaxy as a function of the impact parameter of the incident sight-line with respect to the centre of the gaseous halo.  In this paper, we have tested the TC08 model using a subset of quasar pairs (both lensed and physical), triples, and quads with intervening \mgii\ absorption in one or all of the observed sight-lines.  The mean redshift in our sample was $\langle z_{abs} \rangle=1.099$, covering a range of $z=0.043 - 2.066$.

Using a statistical approach, we tested whether the TC08 model can reproduce the observed variation in \mgii\ absorption strength between paired sight-lines.  We varied the values of two model parameters: characteristic length $\ell_c$ and covering fraction $f_c$.  As two sight-lines become arbitrarily close on the sky, we expect there to be some small separation below which the difference in \mgii\ absorption between sight-lines is negligible; we parameterize this effect as the characteristic length.  The covering fraction measures the percentage of the gaseous halo projected on the sky that will produce \mgii\ absorption.

The best fit parameters were $\ell_c=0.5~h^{-1}_{70}$~kpc and $f_c=0.6$ indicating that it is common to have large differences in magnitude of \mgii\ absorption between sight-lines that are separated by proper distances less than 0.5$~h^{-1}_{70}$ kpc.  The covering fraction we find is inconsistent with that found by \nocite{CT08}{Chen} \& {Tinker} (2008) at 80-86\% if $\ell_c>2~h^{-1}_{70}$\,kpc.  Our best-fit covering fraction is consistent with the ~70\%\ found by \nocite{CH10}{Chen} {et~al.} (2010) at $z<0.5$, and the ~50\%\ found by \nocite{LWP11}{Lundgren} {et~al.} (2011), though the latter measurement made with only one \mgii\ detection.

We can rule out $f_c>0.87$ for all values of $\ell_c$ at 99.7\%.  We can also rule out the region where $\ell_c<1.0~h^{-1}_{70}$ kpc and $f_c<0.3$.  

This study used data from the literature only.  There are a number of quadruply lensed quasars that do not have resolved spectra available for each individual line of sight.  A short search in the literature finds SDSSJ1330+1810 \nocite{OGIBS08}({Oguri} {et~al.} 2008), SDSSJ1251+2935 \nocite{KIOH07}({Kayo} {et~al.} 2007), WFI J2026-4536 and WFI J2033-4723 \nocite{MC04}({Morgan} {et~al.} 2004).  All four objects have either confirmed or suspected intervening \mgii\ absorption.  It would be worth observing these objects spectroscopically with higher spatial resolution in order to separate out the individual lines of sight to see how the \mgii\ absorption varies between sight-lines.

Ultimately, the largest uncertainty in this study is that we are ignorant of the properties and locations of the galaxies that are producing the intervening \mgii\ absorption seen in the spectra of quasar asterisms.  A campaign to locate and study the galaxies responsible for absorption towards quasar asterisms would vastly improve models for \mgii\ absorption, such as the one studied here.

{\it Acknowledgements}.  We would like to thank Sara Ellison, Masamune Oguri, and Chris Churchill  for providing data in support of this work.  PBH and JAR acknowledge funding from NSERC, and JAR from the Faculty of Graduate Studies at York University.




\begin{thebibliography}{}

\bibitem[{Barton} \& {Cooke} 2009]{BC09}
{Barton}, E.~J. \& {Cooke}, J. 2009, \aj, 138, 1817

\bibitem[{Bergeron} \& {Boiss{\'e}} 1991]{BB91}
{Bergeron}, J. \& {Boiss{\'e}}, P. 1991, \aap, 243, 344

\bibitem[{Bordoloi}, {Lilly}, {Knobel}, {Bolzonella},  {Kampczyk}, {Carollo}, {Iovino}, {Zucca}, {Contini}, {Kneib}, {Le Fevre},  {Mainieri}, {Renzini}, {Scodeggio}, {Zamorani}, {Balestra}, {Bardelli},  {Bongiorno}, {Caputi}, {Cucciati}, {de la Torre}, {de Ravel}, {Garilli},  {Kovac}, {Lamareille}, {Le Borgne}, {Le Brun}, {Maier}, {Mignoli}, {Pello},  {Peng}, {Perez Montero}, {Presotto}, {Scarlata}, {Silverman}, {Tanaka},  {Tasca}, {Tresse}, {Vergani}, {Barnes}, {Cappi}, {Cimatti}, {Coppa},  {Diener}, {Franzetti}, {Koekemoer}, {Lopez-Sanjuan}, {McCracken}, {Moresco},  {Nair}, {Oesch}, {Pozzetti}, \& {Welikala} 2011]{BLK11}
{Bordoloi}, R., {Lilly}, S.~J., {Knobel}, C., {Bolzonella}, M., {et al.} 2011, arXiv:1106.0616

\bibitem[{Chen}, {Helsby}, {Gauthier}, {Shectman},  {Thompson}, \& {Tinker} 2010]{CH10}
{Chen}, H., {Helsby}, J.~E., {Gauthier}, J., {Shectman}, S.~A., {Thompson},  I.~B., \& {Tinker}, J.~L. 2010, \apj, 714, 1521

\bibitem[{Chen} \& {Tinker} 2008]{CT08}
{Chen}, H.-W. \& {Tinker}, J.~L. 2008, \apj, 687, 745

\bibitem[{Churchill}, {Kacprzak}, \&  {Steidel} 2005]{CKSiau05}
{Churchill}, C.~W., {Kacprzak}, G.~G., \& {Steidel}, C.~C. 2005, in IAU Colloq.  199: Probing Galaxies through Quasar Absorption Lines, ed. {P.~Williams,  C.-G.~Shu, \& B.~Menard}, 24--41

\bibitem[{Ellison}, {Ibata}, {Pettini}, {Lewis},  {Aracil}, {Petitjean}, \& {Srianand} 2004]{EIP04}
{Ellison}, S.~L., {Ibata}, R., {Pettini}, M., {Lewis}, G.~F., {Aracil}, B.,  {Petitjean}, P., \& {Srianand}, R. 2004, \aap, 414, 79

\bibitem[{Kacprzak}, {Churchill}, {Barton}, \&  {Cooke} 2011]{2011ApJ...733..105K}
{Kacprzak}, G.~G., {Churchill}, C.~W., {Barton}, E.~J., \& {Cooke}, J. 2011,  \apj, 733, 105

\bibitem[{Kacprzak}, {Churchill}, {Steidel}, \&  {Murphy} 2008]{KCS08}
{Kacprzak}, G.~G., {Churchill}, C.~W., {Steidel}, C.~C., \& {Murphy}, M.~T.  2008, \aj, 135, 922

\bibitem[{Kayo}, {Inada}, {Oguri}, {Hall}, {Kochanek},  {Richards}, {Schneider}, {York}, \& {Pan} 2007]{KIOH07}
{Kayo}, I., {Inada}, N., {Oguri}, M., {Hall}, P.~B., {et al.} 2007, \aj,  134, 1515

\bibitem[{Lidman}, {Courbin}, {Kneib}, {Golse},  {Castander}, \& {Soucail} 2000]{LC00}
{Lidman}, C., {Courbin}, F., {Kneib}, J.-P., {Golse}, G., {Castander}, F., \&  {Soucail}, G. 2000, \aap, 364, L62

\bibitem[{L{\'o}pez} \& {Chen} 2011]{LC11}
{L{\'o}pez}, G. \& {Chen}, H.-W. 2011, arXiv:1110.3321

\bibitem[{Lopez}, {Hagen}, \& {Reimers} 2000]{LH00}
{Lopez}, S., {Hagen}, H.-J., \& {Reimers}, D. 2000, \aap, 357, 37

\bibitem[{Lovegrove} \& {Simcoe} 2011]{LS11}
{Lovegrove}, E. \& {Simcoe}, R.~A. 2011, \apj, 740, 30

\bibitem[{Lundgren}, {Wake}, {Padmanabhan}, {Coil},  \& {York} 2011]{LWP11}
{Lundgren}, B.~F., {Wake}, D.~A., {Padmanabhan}, N., {Coil}, A., \& {York},  D.~G. 2011, \mnras, 417, 304

\bibitem[{Maller} \& {Bullock} 2004]{MB04}
{Maller}, A.~H. \& {Bullock}, J.~S. 2004, \mnras, 355, 694

\bibitem[{M{\'e}nard}, {Wild}, {Nestor}, {Quider},  \& {Zibetti} 2009]{MW09}
{M{\'e}nard}, B., {Wild}, V., {Nestor}, D., {Quider}, A., \& {Zibetti}, S.  2009, \mnras, submitted, arXiv:0912.3263

\bibitem[{Morgan}, {Caldwell}, {Schechter}, {Dressler},  {Egami}, \& {Rix} 2004]{MC04}
{Morgan}, N.~D., {Caldwell}, J.~A.~R., {Schechter}, P.~L., {Dressler}, A.,  {Egami}, E., \& {Rix}, H. 2004, \aj, 127, 2617

\bibitem[{Oguri}, {Inada}, {Blackburne}, {Shin}, {Kayo},  {Strauss}, {Schneider}, \& {York} 2008]{OGIBS08}
{Oguri}, M., {Inada}, N., {Blackburne}, J.~A., {Shin}, M., {Kayo}, I.,  {Strauss}, M.~A., {Schneider}, D.~P., \& {York}, D.~G. 2008, \mnras, 391,  1973

\bibitem[Press, Teukolsky, Vetterling, \&  Flannery 2007]{numRec07}
Press, W.~H., Teukolsky, S.~A., Vetterling, W.~T., \& Flannery, B.~P. 2007,  Numerical Recipes 3rd Edition: The Art of Scientific Computing, 3rd edn. (New  York, NY, USA: Cambridge University Press)

\bibitem[{Rauch}, {Sargent}, \& {Barlow} 1999]{RS99}
{Rauch}, M., {Sargent}, W.~L.~W., \& {Barlow}, T.~A. 1999, \apj, 515, 500

\bibitem[{Spergel}, {Bean}, {Dor{\'e}}, {Nolta},  {Bennett}, {Dunkley}, {Hinshaw}, {Jarosik}, {Komatsu}, {Page}, {Peiris},  {Verde}, {Halpern}, {Hill}, {Kogut}, {Limon}, {Meyer}, {Odegard}, {Tucker},  {Weiland}, {Wollack}, \& {Wright} 2007]{wmap3}
{Spergel}, D.~N., {Bean}, R., {Dor{\'e}}, O., {Nolta}, M.~R., {et al.} 2007, \apjs, 170, 377

\bibitem[{Steidel}, {Dickinson}, \& {Persson} 1994]{SDP94}
{Steidel}, C.~C., {Dickinson}, M., \& {Persson}, S.~E. 1994, \apjl, 437, L75

\bibitem[{Steidel}, {Erb}, {Shapley}, {Pettini},  {Reddy}, {Bogosavljevi{\'c}}, {Rudie}, \& {Rakic} 2010]{SSP10}
{Steidel}, C.~C., {Erb}, D.~K., {Shapley}, A.~E., {Pettini}, M., {Reddy}, N.,  {Bogosavljevi{\'c}}, M., {Rudie}, G.~C., \& {Rakic}, O. 2010, \apj, 717, 289

\bibitem[{Tinker} \& {Chen} 2008]{TC08}
{Tinker}, J.~L. \& {Chen}, H. 2008, \apj, 679, 1218

\bibitem[{Warren}, {Abazajian}, {Holz}, \&  {Teodoro} 2006]{WAHL06}
{Warren}, M.~S., {Abazajian}, K., {Holz}, D.~E., \& {Teodoro}, L. 2006, \apj,  646, 881

\bibitem[{Zibetti}, {M{\'e}nard}, {Nestor}, {Quider},  {Rao}, \& {Turnshek} 2007]{ZMNQ07}
{Zibetti}, S., {M{\'e}nard}, B., {Nestor}, D.~B., {Quider}, A.~M., {Rao},  S.~M., \& {Turnshek}, D.~A. 2007, \apj, 658, 161

\end{thebibliography}
\end{document}